\documentclass[10pt,twocolumn,letterpaper]{article}
\usepackage{usenix, times,color,graphicx,amsmath,array,booktabs, amssymb, subcaption}
\usepackage{bcprules}
\usepackage[normalem]{ulem}
\usepackage{xcolor}
\usepackage{multirow}
\usepackage{colortbl}
\usepackage{comment}
\usepackage{algorithmicx}
\usepackage[ruled, vlined, linesnumbered]{algorithm2e}
\usepackage{url}
\usepackage{amsmath}
\usepackage{wrapfig}
\usepackage{amsfonts}
\usepackage{amsthm}
\usepackage{xspace}
\usepackage{tightenum}
\usepackage{wrapfig}
\usepackage{adjustbox}
\usepackage{microtype}
\usepackage{multirow}
\usepackage{tabulary}
\usepackage{floatflt}
\usepackage{mdframed} 
\usepackage{textcomp}
\usepackage[small,compact]{titlesec}
\usepackage{enumitem}
\usepackage[normalem]{ulem}
\usepackage{hyperref}
\usepackage[T1]{fontenc}
\usepackage{tikz}
\usepackage{mdframed}
\usepackage[subtle]{savetrees}
\usepackage{svg}
\usepackage[font=small]{caption}
\usepackage{tabularx}
\usepackage{makecell}
\usepackage[available,functional,reproduced]{usenixbadges}

\usepackage{listings, listings-rust}

\mdfdefinestyle{MyFrame}{%
    linecolor=black,
    outerlinewidth=2pt,
    innertopmargin=4pt,
    innerbottommargin=4pt,
    innerrightmargin=4pt,
    innerleftmargin=4pt,
        leftmargin = 4pt,
        rightmargin = 4pt,
    backgroundcolor=gray!40
}

\hypersetup{
  colorlinks,
  linkcolor={red!50!black},
  citecolor={blue!50!black},
  urlcolor={blue!80!black}
}
\newcommand{\squishlist}{
   \begin{list}{$\bullet$}
    { \setlength{\itemsep}{0pt}      \setlength{\parsep}{3pt}
      \setlength{\topsep}{3pt}       \setlength{\partopsep}{0pt}
      \setlength{\leftmargin}{1.0em} \setlength{\labelwidth}{1em}
      \setlength{\labelsep}{0.5em} } }
\newcommand{\squishend}{
    \end{list}  }

\newcommand{\tool}[0]{DRust\xspace}

\definecolor{codegreen}{rgb}{0,0.5,0}
\definecolor{codeblue}{rgb}{0.1,0.1,0.9}
\definecolor{codegray}{rgb}{0.5,0.5,0.5}
\definecolor{codepurple}{rgb}{0.58,0,0.82}
\definecolor{backcolour}{rgb}{0.95,0.95,0.92}
\definecolor{codeblack}{RGB}{38, 50, 56}
\definecolor{codecomment}{RGB}{111, 129, 147}

\definecolor{lightbg}{RGB}{250, 250, 250}

\lstdefinelanguage{Rust}{
  keywords={impl, fn, let, mut, match, if, else, while, for, loop, pub, use, struct, enum},
  keywordstyle=\color{codeblue}\bfseries,
  identifierstyle=\color{black},
  sensitive=true,
  comment=[l]{//},
  morecomment=[s]{/*}{*/},
  commentstyle=\color{codegreen}\itshape,
  stringstyle=\color{orange},
  morestring=[b]",
  escapeinside={/*@}{*/},
}

\lstdefinestyle{mystyle}{
    numberstyle=\tiny\color{black},
    basicstyle=\ttfamily\scriptsize,
    captionpos=b,                    
    numbers=left,                    
    numbersep=5pt,                  
    xleftmargin=1.7em,
}
\newcommand{\us}[0]{$\mu$s\xspace}

\newcommand{\codeIn}[1]{{\small\texttt{#1}}}

\newenvironment{myquote}%
  {\list{}{\leftmargin=0.1in\rightmargin=0.1in}\item[]}%
  {\endlist}

\newcommand{\MyPara}[1]{\vspace{.1em}\noindent\textit{\textbf{#1}}}

\newcommand{\lab}{\mbox {$K$}} %labeling
\newtheorem{definition}{Definition}[section]
\newtheorem{lemma}{Lemma}[section]

\newcommand{\squishlistree}{
   \begin{list}{$\bullet$}
    { \setlength{\itemsep}{0pt}      \setlength{\parsep}{0pt}
      \setlength{\topsep}{3pt}       \setlength{\partopsep}{0pt}
      \setlength{\leftmargin}{1em} \setlength{\labelwidth}{1em}
      \setlength{\labelsep}{0.5em} } }

\newcommand{\squishlisttwo}{
   \begin{list}{$\bullet$}
    { \setlength{\itemsep}{0pt}    \setlength{\parsep}{0pt}
      \setlength{\topsep}{0pt}     \setlength{\partopsep}{0pt}
      \setlength{\leftmargin}{2em} \setlength{\labelwidth}{1.5em}   
      \setlength{\labelsep}{0.5em} } }

\definecolor{ForestGreen}{RGB}{34,139,34}

\newcommand{\eg}{\hbox{\emph{e.g.}}\xspace}
\newcommand{\ie}{\hbox{\emph{i.e.}}\xspace}

\newcommand{\etc}{\hbox{\emph{etc.}}\xspace}

\newcommand{\ignore}[1]{}

\makeatletter
\newcommand*{\circled}{\@ifstar\circledstar\circlednostar}
\makeatother
 
\newcommand*\circledstar[1]{%
   \tikz[baseline=(C.base)]
     \node[%
       fill=black!20,
       circle,
       minimum size=1em,
       text=black,
       font=\footnotesize,
       inner sep=0.3pt
     ](C) {#1};%
}
\newcommand*\circlednostar[1]{%
   \tikz[baseline=(C.base) - .6em]
     \node[%
       fill=black,
       text=white,
       circle,
       minimum size=.8em,
       font={\bf \footnotesize},
       inner sep=0.2pt
     ](C) {#1};%
}

\newcommand\blfootnote[1]{%
  \begingroup
  \renewcommand\thefootnote{}\footnote{#1}%
  \addtocounter{footnote}{-1}%
  \endgroup
}

\begin{document}

\interfootnotelinepenalty 100000
\widowpenalty 100000
\clubpenalty 100000
\newfont{\tf}{phvro at 9.5pt}
\newfont{\tft}{phvro at 7.25pt}

% \twocolumn[\begin{@twocolumnfalse}

% \begin{centering}
% \vspace{0.1cm}
% {\large \bf DRust: Language-Guided Distributed Shared Memory with Fine Granularity, Full Transparency, and Ultra Efficiency\\}
% \vspace{0.1cm}
% OSDI'24 Submission \#77 
% \vspace{-0.3cm}

% \end{centering}

% \vspace{\baselineskip}

% \end{@twocolumnfalse}]

\title{DRust: Language-Guided Distributed Shared Memory with Fine Granularity,\\Full Transparency, and Ultra Efficiency}
\author{\rm{Haoran Ma}$^{\dag\star}$\hspace{1.4em}Yifan Qiao$^{\dag}$\hspace{1.4em}Shi Liu$^{\dag\star}$\hspace{1.4em}Shan Yu$^{\dag}$\hspace{1.4em}Yuanjiang Ni$^{\psi}$\hspace{1.4em}Qingda Lu$^{\psi}$\hspace{1.4em}Jiesheng Wu$^{\psi}$\hspace{1.4em}\\[.3em]\rm{Yiying Zhang}$^{\ddag}$\hspace{1.4em}Miryung Kim$^{\dag}$\hspace{1.4em}Harry Xu$^{\dag}$
\\[.8em]
UCLA$^{\dag}$\hspace{2.5em}UCSD$^{\ddag}$\hspace{2.5em}Alibaba Group$^{\psi}$ \\[1em]
}
\maketitle

\setcounter{page}{1}

\begin{abstract}
Despite being a powerful concept, distributed shared memory (DSM) has not been made practical due to the extensive synchronization needed between servers to implement memory coherence. This paper shows a practical DSM implementation based on the insight that the ownership model embedded in programming languages such as Rust automatically constrains the order of read and write, providing opportunities for significantly simplifying the coherence implementation if the ownership semantics can be exposed to and leveraged by the runtime. This paper discusses the design and implementation of \tool, a Rust-based DSM system that outperforms the two state-of-the-art DSM systems GAM and Grappa by up to \textbf{2.64$\times$} and \textbf{29.16$\times$} in throughput, and scales much better with the number of servers. 
\end{abstract}
\blfootnote{$^\star$ Part of the work was done when Haoran Ma and Shi Liu interned at Alibaba Group.}
\section{Introduction}
\label{sec:intro}

The concept of distributed shared memory (DSM) received significant attention during the early years of distributed computing systems. This era witnessed a plethora of pioneering efforts, as exemplified by seminal works such as~\cite{li1988ivy, li1989memory,carter1991implementation,carter1995techniques,stets1997cashmere, bennett1990munin, minnich1993mether, fleisch1987distributed, campbell1987choices,nieplocha1994global,nieplocha1996global,nieplocha2002combining,gustavson1992scalable}.  
DSM offers the power of parallel computing using multiple processors and machines and, more crucially, streamlines the development of distributed applications with a unified, contiguous memory view. 

The initial enthusiasm for DSM was tempered by significant performance bottlenecks, primarily due to the low network speeds prevalent during its nascent stages.
%These challenges hindered the widespread adoption of DSM.\miryung{is this only performance issue or programmability issue? since we are proposing a language based solution.}
Recent advances in hardware and networking technologies~\cite{firebox-keynote-fast14, shoal-nsdi19, farm-nsdi14, ccixconsortium, genzconsortium, intel-fast, mellanox-connectx-6, opencapi-18, infiniband-verbs-performance, disaggregated_mem@hotnets13, pim-enabled-instructions-isca15, pardis-isca12, the-machine-ross15, the-machine-hp, memory-disaggregation-isca09, ramcloud@tocs15, cxl} have revitalized the DSM explorations. Several new DSM systems~\cite{cai2018efficient, nelson2015latency, taranov2021corm, wang2021concordia, kaxiras2015turning, shan2017distributed} were proposed in recent years to take advantage of these enhanced networks. 
%Employing remote direct memory access (RDMA) for data transmission and synchronization between servers, these systems yield better performance compared to the DSM of earlier days. 
However, these systems are still far from achieving satisfactory performance, exhibiting poor scalability and substantial slowdown compared to their single-machine counterparts. This is mainly due to the intensive synchronization operations needed to ensure memory coherence across servers.  %\miryung{we need a connector here that a language based soluton can ease the difficulty of ensuring memory consistency, which can aid the performance problem.}

\MyPara{State of the art.} The majority of existing DSM systems~\cite{amza1996treadmarks,kaxiras2015turning,cai2018efficient,wang2021concordia} adopt an approach to achieve data consistency by adhering to the following invariant: for each data block to be accessed, the block is either located on a single node with potential read and write access, or it is replicated across multiple nodes with each having read access only. Prior to a server attempting to access a block, a DSM system checks the state of the block, invalidates copies of that block on all other servers, and then transmits the block to the requesting server. This synchronization process necessitates multiple network round trips. Even with RDMA, the incurred latency is still orders of magnitude higher compared to a single local access, significantly degrading overall performance. Effectively reducing the number of synchronizations is, therefore, crucial for minimizing DSM overhead and rendering it feasible for real-world deployment.

A practical strategy to minimize synchronization overhead involves implementing high-level protocols to guarantee exclusive access for each server. For instance, Apache Spark~\cite{spark} utilizes an immutable data structure known as a resilient distributed dataset (RDD) for distributed access. However, RDD only facilitates coarse-grained distributed access, limiting each server to accessing a distinct partition of an RDD. While increasing access granularity enhances performance, it comes at the expense of reduced generality\textemdash Spark is tailored for bulk processing of batch data and is incapable of supporting distributed applications requiring object-level accesses, such as social networks where objects of various types and sizes (\eg, images, connections, \etc) are created and manipulated upon each user request. %\qd{maybe some more explanations with fine-grain object examples in social network applications?}

%Another example is found in recent research~\cite{semeru@osdi2020, ruan2020aifm} that facilitates computation offloading to remote memory. Although these techniques can schedule specific functions onto remote servers, they prohibit concurrent accesses of the same data structure on different servers, thereby eliminating the need for synchronization. However, these techniques represent simplified versions of DSM without its full capability to support concurrent fine-grained data modifications. The goal of this work is to create a full-fledged DSM system that supports fine-grained data access with exceptional efficiency. \yq{Can we skip this para in the submission version? We can mention them in related work.}

\MyPara{Insights.} Our main observation is that synchronization overheads in existing DSM systems are introduced primarily due to the use of a generic approach that overlooks semantic information from programs. For example, 
many real-world concurrent programs are engineered with a single-writer-multiple-reader (SWMR) discipline to ensure correctness during concurrent operations. Leveraging such information can potentially eliminate the need to check the state of remote data blocks before accessing them, leading to dramatically improved performance. A major challenge is, however, how to expose such semantics in a sensible way so that the DSM system can see and act upon it.

%However, the challenge is how to establish a connection between the program and the underlying DSM system, and how the DSM system can efficiently leverage this semantic information to enhance its performance. \miryung{this part needs more crystalized argument, what is the nature of connection abstraction between a program and the DSM, why a language based approach is needed?} Addressing this challenge necessitates a novel approach to bridge the program’s operational logic with the system's functional dynamics. \miryung{what is the operatinoal logic and what is functional dynamics?}

One approach to convey such semantics, as demonstrated by AIFM~\cite{ruan2020aifm} and Midas~\cite{qiao2024midas}, involves exposing APIs that developers can invoke to specify program regions accessible only by a single writer. However, this process is cumbersome and error-prone, demanding a profound understanding of potential executions and involving substantial program writing. Our key insight in this endeavor is that the SWMR programming paradigm aligns seamlessly with \emph{ownership types}, which have already been integrated into programming languages like Rust~\cite{rustweb}. Rust is widely employed in the system community for dependable and secure implementation of low-level systems code.

Rust's ownership type inherently upholds SWMR properties in any compiled Rust program. The fundamental concept behind the ownership type is that each value is ensured to have a single unique variable as its owner throughout the execution. While multiple references to a value are allowed, only the owner and mutable references can modify the value. Moreover, only one of these references is permitted to be used for modifying the value at any given point.

When developing a DSM system on top of an ownership-based language like Rust, SWMR semantics are inherently embedded in any Rust program \emph{by design}. Effortlessly extracting such information becomes possible with basic compiler support, sparing developers from the need for code rewriting. Utilizing the SWMR semantics from the program leads to a considerably simplified process for accessing data in DSM. In the case of a write access, the ownership type ensures exclusive access to the data. Consequently, \tool can move the data to the requesting machine, performing the write there without explicitly invalidating its copies on other machines. In the case of a read access,  data can be efficiently replicated to (and cached in) each requesting machine, benefiting from the compiler-provided assurance of freedom from concurrent writes.

%This model inherently provides the exact information needed by an efficient data handling paradigm in DSM. \miryung{if is a direct adoption of available Rust feature, why is it novel in this system implementation context? we may want to say some challenges associated with implementing rust style semantics in distributed context such as data movement.} With the ownership model, rather than engaging in complex synchronization across various machines, 

This paper presents \tool, an efficient Rust-based DSM implementation that enables object-level concurrent accesses by leveraging the SWMR semantics made explicit by Rust's ownership type. \tool automatically turns a single-machine Rust program into a DSM-based distributed version \emph{without requiring code rewriting}. 
%For further performance optimizations, \tool provides \emph{affinity annotations} that one can easily use to annotate computations or data structures, providing locality hints for the runtime to make informed scheduling decisions. 
While extracting the ownership semantics appears straightforward, leveraging it to implement a distributed coherence protocol correctly and efficiently presents two main challenges.

The first challenge is \emph{how to manage memory correctly and efficiently.} Rust's ownership type system is inherently designed for a single-machine environment, where the memory address of an object remains constant post-creation. This assumption is disrupted in a distributed environment, where objects may be migrated or duplicated on different machines. Such actions can lead to the risk of dangling pointers, potentially breaking memory coherence.

To tackle these issues, \tool builds a global heap spanning multiple servers based on the idea of partitioned global address space~\cite{coarfa2005evaluation}. Each object in the heap has a unique global address in the address space, which can be used for accessing the object from any server. \tool re-implements Rust's memory management constructs to allocate objects in the global heap. Given that a server can have cached objects (to accelerate reads), \tool carefully crafts an ownership-based cache coherence protocol upon the global heap abstraction to achieve both memory coherence and efficiency (\S\ref{sec:design:pa:memory-management}). 

In a nutshell, our coherence protocol leverages the ownership semantics to eliminate the need for explicit cache invalidation. It allows multiple readers to fetch a copy of the object from its host server and cache it, but disallows any change to the global address and the value of the object. When a write access occurs, it must first borrow the ownership, at which point \tool moves the object in the global heap to a new address on the server issuing the write. The address change of the object automatically invalidates cache copies that use the stale address and triggers the subsequent readers to update the cache by fetching the object from its latest address.

The second challenge is \emph{how to support transparency in programming.} Rust's standard libraries and programs were originally built for running on a single machine, and they cannot deal with distributed resources in a cluster. For example, a Rust program running on server A cannot spawn a thread on another server B, let alone synchronize threads between A and B. To enable a Rust program to run \emph{as is} under \tool, we provide distributed threading utilities by restructuring critical elements of the Rust standard library, including threading, communication channels, and shared-state locks (\S\ref{sec:design:pa:std-lib-support}). Our adapted libraries offer the same interfaces, making them compatible with single-machine Rust programs, but internally invoke our distributed scheduler, which determines where to run the thread and facilitates cross-server synchronization. We built them atop the ownership-based memory model, enabling the \tool runtime to safely pass references of objects between threads and automatically fetch the value from the global heap upon dereferencing. 

With our programming abstractions, a Rust application can start on a single server and gradually spawn its threads to other servers. Under the hood, \tool employs a runtime to manage distributed physical compute and memory resources for the application. The runtime runs as a process on each node in the cluster, and they work cooperatively for cross-server memory allocation and thread scheduling. The runtime prioritizes the current server for object allocation and thread creation, but it will schedule the resource allocation request to another server under memory pressure (\S\ref{sec:rtsys:apprt}). To make cluster-wise decisions such as deciding the target server for global memory allocation and thread creation, \tool has a global controller that is launched together with the application. The global controller communicates with \tool runtime on each node to collect resource usage information and applies adaptive policies to achieve load balance (\S\ref{sec:rtsys:controller}).

\MyPara{Results.} We evaluated our system on four real-world applications in an eight-node cluster. Our evaluation demonstrated an average of 2.02$\times$ and 9.48$\times$ (up to 2.64$\times$ and 29.16$\times$) speedup compared with two state-of-the-art DSM systems GAM and Grappa, respectively.
Furthermore, \tool incurred a mere 2.42\% slowdown compared to the original Rust program on a single machine with sufficient resources. \tool is available at \url{https://github.com/uclasystem/DRust}.
\section{Background in Ownership}
\label{sec:background}

Over the past decades, numerous programming languages have been designed to provide safe memory management and data sharing. At the core of such a design is often a tradeoff between memory abstraction level and management efficiency. The ownership concept, and the Rust programming language built upon, are considered promising solutions that achieve a sweet spot between abstraction and efficiency. This section provides an overview of these techniques and explains how ownership can benefit DSM implementations.

\MyPara{Ownership Type.}
The ownership model has a long history in pursuit of memory-safe language designs and type systems~\cite{linearlisp1992, vault2001, cyclone2002, ynot2008, crust2015, mezzo2016}. It has also inspired many systems for safe and efficient resource management~\cite{singularity2007, redleaf2020, theseus2020, rayownership2021}. At a high level, ownership enhances a language's type system in a way that guarantees the memory and thread safety of a program with type checking done at compile time. The ownership model encompasses a range of concepts, among which the most important are \emph{lifetimes} and \emph{borrowing}.

An ownership-based type system uses lifetimes to control the allocation/deallocation of objects. It enforces that each object must have one and only one owner at a time. This allows the compiler to statically track an object's lifetime via its owner, and immediately deallocate the object once its owner goes out of scope, preventing memory leaks without using garbage collection that can introduce disruptive pauses to program execution.

To access an object, a program can create a reference from its owner, but the reference must ``borrow'' the permission from the owner, and ``return'' it to the owner after the access. Specifically, the type system allows the creation of multiple \emph{immutable references} to an object from its owner for concurrent reads but prohibits any write with these references. It allows only one mutable reference to the object only when no other (mutable or immutable) references exist. Through borrowing, the ownership type disallows simultaneous writers and hence prevents data races. In addition, references must return the borrowed permission when they go out of scope. For any program that demonstrates type soundness, the type checker guarantees that references to an object can only reside within the object's lifetime; the object can be safely and automatically deallocated when its owner goes out of scope, by which time it has already lost all its references.

Finally, ownership can be transferred from one owner to another\textemdash\eg, at a function call, the creation of a thread, or message passing (\ie, via \codeIn{channel}). However, the type system enforces that ownership transfer must occur in the absence of ``borrowing''. In other words, no other references can exist in scope when transferring the ownership,  preventing data races during ownership transfers.

The guarantees provided by the ownership model with respect to object lifetime and data sharing can be summarized with the following four invariants:
\begin{enumerate}[nosep, left=0pt, itemsep=0pt, topsep=0pt]
    \item \textbf{Singular Owner}: each value has one single owner at any time (which must also belong to one single thread).\label{inv:1}
    \item \textbf{Safe Borrowing}: All references are created from the owner; permission borrowing and returning guarantees that references that can be used to access the object must be valid.\label{inv:2}
    \item \textbf{Single Writer}: Each object allows one mutable reference at most, and it cannot coexist with any other references in the same scope.\label{inv:3}
    \item \textbf{Multiple Reader}: Multiple references are permitted only when all of them are immutable.\label{inv:4}
\end{enumerate}
The last two invariants are commonly called the single-writer-multiple-reader (SWMR) property in the DSM literature~\cite{prime-coherence-book2020}. 
%These semantics, if leveraged properly, can greatly benefit distributed shared memory design by simplifying memory management and the cache coherence protocol.

\begin{figure}[t]
    \centering
    \includegraphics[width=\linewidth]{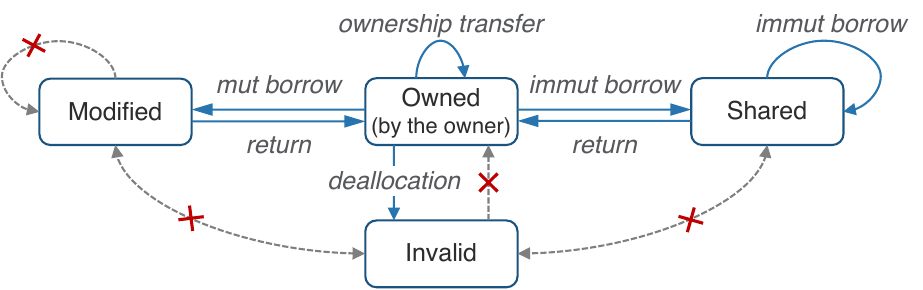}
    \vspace{-1.5em}
    \caption{State machine for Rust's ownership-based memory model.
    }
    \label{fig:cache-states}
  \vspace{-1.5em}
\end{figure}

\MyPara{Rust Language.}
Rust offers a practical implementation of ownership and is designed with a range of zero-cost abstractions for efficient fine-grained resource management. Figure~\ref{fig:cache-states} depicts the state machine for Rust's ownership-based memory model. At a high level, this model restricts that the owner is always in the O (owned) state, and transitions between M (modified), S (shared), and I (invalid) must go through the O state\footnote{A transition from M to S is also possible as an optimization in Rust.}. Clearly, a distributed implementation of this approach avoids broadcasts or snooping, and only requires peer-to-peer message passing.

\begin{lstlisting}[language=Rust, float=t, floatplacement=t, 
caption={A simple accumulator implementation in Rust.},
belowskip=-2.5em,
label={lst:accumulator-sample}]
pub struct Accumulator { pub val: Box<i32>, } /*@ \label{lst:simple-acc:acc-stt} */
impl Accumulator {
  pub fn add(&mut self, delta: &i32)->i32 {
    *self.val += *delta;
    *self.val
  }
} /*@ \label{lst:simple-acc:acc-end} */
fn main() {
  // Allocates two integers in the heap.
  let val: Box<i32> = Box::new(5); // val is an owner./*@\label{lst:simple-acc:val-box}*/
  let mut b: Box<i32> = Box::new(0); // b is an owner./*@\label{lst:simple-acc:b-box}*/
  // Ownership is transferred from val to a.val
  let mut a = Accumulator{val}; /*@ \label{lst:simple-acc:inst-acc} */
  { // Only one mutable reference is allowed. /*@ \label{lst:simple-acc:mut-stt} */
    let mutr: &mut i32 = &mut *b;
    // No other reference is allowed now.
    /* let another_r = &*b; */ // COMPILE ERROR!
    *mutr = 10; // b == 10
  } /*@ \label{lst:simple-acc:mut-end} */
  { // Multiple immutable references are allowed./*@\label{lst:simple-acc:immut-stt}*/
    let (b_r1, b_r2): (&i32, &i32) = (&*b, &*b); 
    // Mutable reference is prohibited now.
    /* let b_mutr = &mut *b; */ // COMPILE ERROR!
    // Passing by references won't transfer ownership.
    let sync_add = a.add(b_r1); // a.val == 15
    let sync_add = a.add(b_r2); // a.val == 25
  } /*@ \label{lst:simple-acc:immut-end} */
  {// Ownership of a and b is moved to the new thread./*@\label{lst:simple-acc:conc-stt}*/
    // No reference should or can borrow a or b now.
    let async_add = thread::spawn(move || 
      a.add(&*b) // a.val == 35
    ).join(); // lifetime of a and b ends /*@ \label{lst:simple-acc:join} */
    // Current thread cannot access a and b anymore.
    /* println!("{}", a.val); */ // COMPILE ERROR!
  } /*@ \label{lst:simple-acc:conc-end} */
}
\end{lstlisting}

Listing~\ref{lst:accumulator-sample} exemplifies a simple accumulator implemented in Rust (Lines \ref{lst:simple-acc:acc-stt}--\ref{lst:simple-acc:acc-end}). The \codeIn{Accumulator} struct keeps an integer \codeIn{val} and exposes an interface \codeIn{add} to increment the value. Rust uses a smart pointer type \codeIn{Box<T>} to store values on the heap; this pointer serves as the initial owner of the referenced value, as shown in Line \ref{lst:simple-acc:val-box} and \ref{lst:simple-acc:b-box}. Line \ref{lst:simple-acc:inst-acc} instantiates \codeIn{Accumulator a}, where the ownership is implicitly transferred from \codeIn{val} to \codeIn{a.val} during its initialization. Rust allows the creation of mutable and immutable references to access the value. For example, Lines \ref{lst:simple-acc:mut-stt}--\ref{lst:simple-acc:mut-end} create a singular mutable reference (\codeIn{\&mut}) to \codeIn{b} and set its value to 10. Similarly, Lines \ref{lst:simple-acc:immut-stt}--\ref{lst:simple-acc:immut-end} create two immutable references (\codeIn{\&}) to \codeIn{b} and add them to \codeIn{a} via two function calls. Note that passing references as arguments in function calls does not transfer their ownership. 

Finally, Rust allows spawning new threads for concurrent programming, as shown in Lines \ref{lst:simple-acc:conc-stt}--\ref{lst:simple-acc:conc-end}. A new thread is created via \codeIn{thread::spawn}, where the use of \codeIn{move} captures \codeIn{a} and \codeIn{b} in the current scope and transfers their ownership to the newly spawned thread. Rust performs shallow copying for inter-thread communication, where only the pointers stored in \codeIn{a} and \codeIn{b} are transferred to the child thread while the actual values on the heap are not moved. Rust guarantees memory safety of \codeIn{a} and \codeIn{b} by tracking their ownership. At Line \ref{lst:simple-acc:join}, when the child thread finishes its closure (\ie, not necessarily after \codeIn{join}), and \codeIn{a} and \codeIn{b} exit the scope (to which their ownership belongs), their lifetimes terminate and Rust deallocates them from the heap.

\section{Motivation}
\label{sec:motivation}

DSM was proposed to eliminate the barrier of distributed programming by offering the same memory consistency model as single-machine shared memory.
The core of its design is a software-based cache coherence protocol, which mimics a hardware-based approach on multi-core CPUs and synchronizes memory states on different servers by sending control messages between them. 
However, it is notoriously hard to implement cache coherence efficiently at the software level due to the high communication latency between physically disjointed servers.

% \begin{figure}[h!]
%     \centering
%     \includegraphics[scale=.42]{figs/motivation.pdf}
%     \caption{Performance comparison of GAM and pure one-sided RDMA operations. \yq{TODO: update the figure.}\label{fig:motivation}}
% \end{figure}

\MyPara{High Synchronization Overheads for Coherence.}
To gain a high-level understanding of how much improvement can be achieved by improving the cache coherence protocol, we performed an analysis by running a real-world application DataFrame~\cite{polars} with a state-of-the-art DSM system GAM~\cite{cai2018efficient} with a fast network. We first ran Dataframe on a single server with 16 CPU cores and 64GB memory. We then ran it with GAM on eight servers connected by a 40Gbps Infiniband network by evenly distributing the same amount of resources to eight servers (\ie, each server uses 2 CPU cores and 8GB memory). 
Our experiments show a \textbf{2.4$\times$} slowdown when Dataframe runs on eight servers. 
% Figure~\ref{fig:motivation} reports the performance comparison: Dataframe is \textbf{2.4$\times$} slower on the distributed shared memory.
% , indicating that we need at least \yq{Z}$\times$ resources to recover application's performance in a distributed setting even assuming perfect scalability, which is unrealistic in the real world.

A detailed examination reveals that such a slowdown stems primarily from its complicated coherence protocol. GAM runs a directory-based protocol, which assigns each DSM cache block a home node. Upon each object read/write, the home node tracks the state of its cache block and updates all cache copies for the state change, incurring extensive computation and network overhead. We broke down the average time spent on each component when accessing one object in the DSM. Reading a 512-byte (\ie, GAM's default cache block size) uncached object in GAM takes 16\us, 
% among which, tracking cache block states takes \yq{A}\us, network traffic for coherence messages consumes \yq{B}\us, 
while the actual time to read the object over the network is only 3.6\us. In other words, maintaining cache coherence takes \textbf{77\%} of the total time. This large memory access overhead significantly increases operation latency, hindering the practical deployment of distributed shared memory. With the single writer invariant inherent in the ownership model, we expect that most of this overhead can be eliminated, leading to significant ($>2\times$) speedups for each access.

\section{Design}
\label{sec:design}

\begin{figure}[t]
    \centering
    \includegraphics[width=\linewidth]{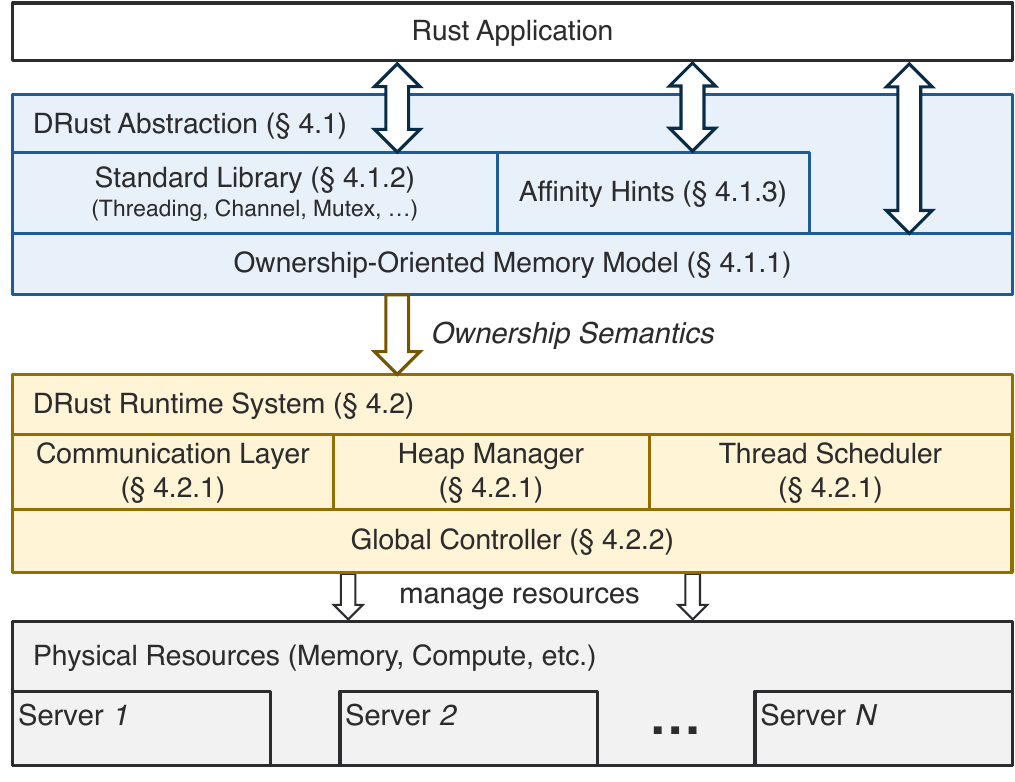}
    \vspace{-1.5em}
    \caption{Design overview of \tool.   \label{fig:overall-design}}
  \vspace{-1em}
\end{figure}

\tool is an efficient DSM framework atop the Rust programming language. 
As shown in Figure~\ref{fig:overall-design}, it consists of Rust-based programming abstractions for DSM (\S\ref{sec:design:abstraction}) and a runtime (\S\ref{sec:design:runtime-system}) that manages distributed physical resources. 

\tool is compatible with standard Rust. 
% A \emph{unmodified} Rust program can be seamlessly re-compiled in \tool and run in a distributed manner. The program (main function) starts on a single server, and gradually scales out by allocating its memory and spawning its threads onto multiple servers. \tool enables the program to freely pass pointers and references of objects between threads running on different servers and dereferencing them anywhere. Objects may be cached on multiple servers and modified during the program execution, and \tool uses a carefully crafted ownership-based coherence protocol to achieve memory consistency with minimal synchronization across servers.
Listing~\ref{lst:dist-accumulator-sample} illustrates how the accumulator (shown in Listing~\ref{lst:accumulator-sample}) runs on \tool distributively without requiring code rewriting.
The program starts running on a single machine A and the \tool runtime gradually allocates its memory and spawns new threads on different machines. 
Specifically, Lines \ref{lst:dist-acc:line:acc-stt}--\ref{lst:dist-acc:line:acc-end} create \codeIn{Accumulator a} and \codeIn{b} where \codeIn{a.val} and \codeIn{b} are in the global heap. We use a global allocator to allocate objects in the global address space and hence these objects may be allocated on a different server. Line \ref{lst:dist-acc:line:local-add} synchronously adds \codeIn{b} to \codeIn{a} by fetching both values \codeIn{a.val} and \codeIn{b} to A's local memory (if they are allocated somewhere else). %Alternatively, 
Line \ref{lst:dist-acc:line:remote-add} spawns a new thread and ships the function closure to perform \codeIn{add} asynchronously. This thread will be scheduled on a different server B if A's compute power has been saturated. In this case, \tool performs shallow copying and only ships the pointers stored in \codeIn{a} and \codeIn{b} to B without actually moving objects in the global heap. The newly-created thread relies on the \tool runtime to detect data locations and fetch objects upon dereferencing.

\begin{lstlisting}[language=Rust, float=t, floatplacement=t, 
caption=\tool seamlessly transforms an unmodified accumulator implemented in Rust into a distributed version.,
belowskip=-2em,
label={lst:dist-accumulator-sample}]
// Unmodified Rust code.
pub struct Accumulator { pub val: Box<i32>, }
impl Accumulator {
  pub fn add(&mut self, delta: &i32)->i32 {
    *self.val += *delta;
    *self.val
  }
}
fn main() {
  // Allocates two integers in the distributed heap. /*@ \label{lst:dist-acc:line:acc-stt} */
  let val: Box<i32> = Box::new(5);
  let b: Box<i32> = Box::new(10);
  let mut a = Accumulator{val}; /*@ \label{lst:dist-acc:line:acc-end} */
  // a.val and b will be fetched to local. 
  let local_add = a.add(&*b); // a.val == 15 /*@ \label{lst:dist-acc:line:local-add} */
  // Only refs to a and b are shipped to remote.
  let remote_add = thread::spawn(move || /*@ \label{lst:dist-acc:line:remote-add} */
    a.add(&*b)).join(); // a.val == 25
}
\end{lstlisting}

\subsection{\tool Programming Abstraction\label{sec:design:abstraction}}
\tool provides each thread with a local stack and abstracts distributed memory as a shared global heap. Each server allocates thread stacks and backs one partition of the global heap with its physical memory. \tool re-implemented core memory management constructs including \codeIn{Box}, \codeIn{\&}, and \codeIn{\&mut} for transparent heap access. This approach hides the complex details of memory allocation/deallocation, moving objects, and coherence maintenance (\S\ref{sec:design:pa:memory-management}). \tool supports distributed threading and synchronization by adapting Rust's standard libraries atop the core language constructs (\S\ref{sec:design:pa:std-lib-support}). Furthermore, \tool offers affinity annotations that allow developers to build more efficient applications by expressing data affinity semantics (\S\ref{sec:design:pa:affinity-annotations}).

\subsubsection{Memory Management \label{sec:design:pa:memory-management}}

\begin{figure}[t]
\centering
\small
\includegraphics[width=\linewidth]{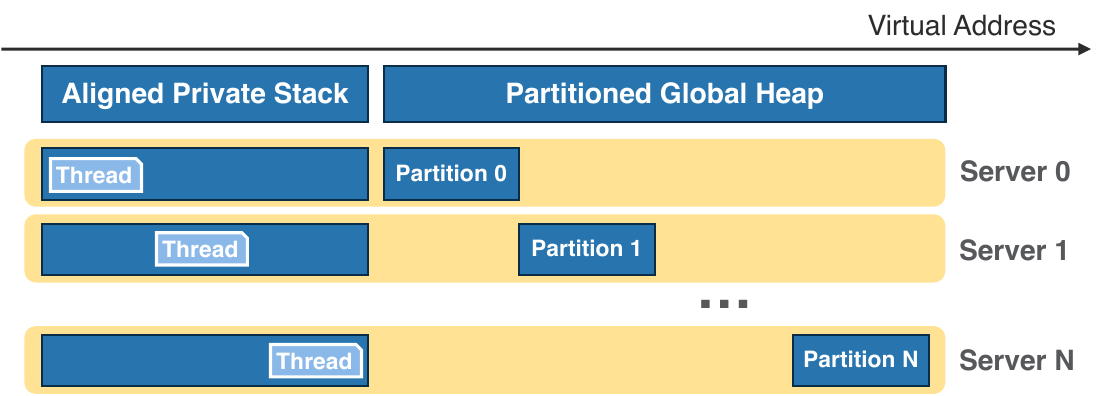}
\vspace{-1em}
\caption{The address space layout of \tool. The stack is private to each thread but they share an aligned address space to ease migration, while the heap is globally shared and partitioned across servers. \label{fig:address-space}}
  \vspace{-1.5em}
\end{figure}

Next, we discuss how \tool (re)implements the memory-related language constructs in Rust to achieve memory safety and memory coherence.

\MyPara{Address Space.}
As shown in Figure~\ref{fig:address-space}, \tool maintains an identical address space layout on all servers. It exposes distributed memory as a coherent shared heap to applications. Embracing the idea of partitioned global address space (PGAS)~\cite{coarfa2005evaluation}, it partitions the heap space and assigns each server a unique address range. The stack, in contrast, is private to each thread. However, \tool aligns the stack space on each server and pads stacks to avoid overlapping. This streamlines thread migration between servers as it allows a thread to keep its private stack address unchanged when being moved.
%., even when it is migrated to another server. 

\MyPara{Coherence Protocol in a Nutshell.}
For efficiency, \tool employs a \emph{call-by-reference} model for newly created threads. Upon creation of a thread, the \tool runtime only passes references or \codeIn{Box} pointers to objects to the newly created thread. Upon dereferencing, objects are fetched to the server where the thread is executed. 

When a \emph{read access} of an object is issued on a server, our runtime simply fetches a copy of the object from its hosting server and places it in its \emph{local cache}. As a result, multiple copies of the same object may exist on different servers. This allows multiple servers to read the object at the same time from their respective cached copies. Fetching a copy of the object for read does not change the object's address in the global space. When a \emph{write access} occurs on an object, the server issuing the write must first
obtain the object's write access permission through a \emph{mutable borrow}. Our reimplementation of mutable borrow (discussed shortly) \emph{moves\footnote{The term ``copy'' is used to describe the process of adding an object into the cache without changing its global address. The term ``move'' means relocating the object into a server's heap partition, which requires changing its global address.} the object in the global heap to a new address} that belongs to that server. In doing so, the object's cached copies on other servers are automatically invalidated without sending explicit invalidation messages\textemdash subsequent reads on these servers must obtain an immutable reference to the object through an immutable borrow from its owner pointer, which has been updated to the new address immediately after the mutable borrow returns. Upon identifying the owner's address change, each immutable borrow would direct a server to fetch a fresh version of the object from the new address as opposed to relying on a stale copy residing in its cache. 

Note that this is a general protocol that covers the case that the object is on the same server that issues the write\textemdash as long as the server moves the object into a different location in the global heap, no other servers can read the stale copies of the object. 
However, this is not efficient as each local write requires moving the object to a new address. 
% \tool implements an optimization (discussed at the end of this subsection) that efficiently deals with local writes.
To address this inefficiency, \tool employs a pointer-coloring technique, inspired by the designs of many concurrent garbage collectors \cite{zgc,mako-pldi22}. Discussed at the end of this subsection, this technique offers a more efficient solution for handling local writes.
% Inspired by concurrent garbage collectors~\cite{zgc,mako-pldi22}, \tool optimizes local writes via pointer coloring that avoids unnecessary object movements on the same server (discussed next).

\begin{figure}[t]
    \centering
    % \small
    \includegraphics[width=\linewidth]{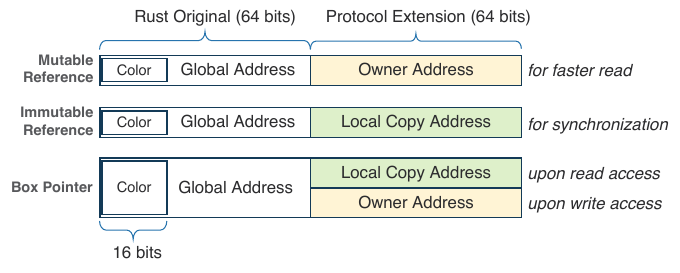}
  \vspace{-2em}
    \caption{\tool repurposes Rust pointers and references to contain a global heap address and an extension field for its coherence protocol.} %\yq{Do we want to annotate "upon read/write access" similar to them in the previous figure?}
    \label{fig:pointer-layout}
  \vspace{-1em}
\end{figure}

% Figure~\ref{fig:pointer-layout} depicts the representation of \tool's \codeIn{Box} pointer and references. Unlike in Rust where addresses are always local, they store a global heap address that potentially points to the heap partition on another server. Additionally, \tool adds a 64-bit extension field for its coherence protocol, as elaborated below.
\MyPara{Pointer Layout.} In order to support this protocol, each pointer must remember not only the object's global address, but also the address of the cached copy in a server's local cache (to avoid redundant remote fetches). As such, we modify Rust's pointer structure, as illustrated in Figure~\ref{fig:pointer-layout}. \tool internally extends each Rust \codeIn{Box} pointer and reference with an additional 64-bit field, which is used differently for read and write access. At a high level, the field records the address of the cached copy for faster read accesses; for write accesses, this field records the address of the object's owner for post-write synchronization. Additionally, \tool reserves the highest 16 bits in the global address field as ``color'' bits. These bits record the version number of the pointer and play a crucial role in \tool's efficient handling of local writes.
% \revise{Additionally, \tool reserves the highest 16 bits in the global address field as ``color'' bits. These bits record the version number of the pointer, and they are incremented upon mutable borrow returns. By comparing the version numbers, subsequent immutable borrows can still identify the owner's address change even though the object is only accessed locally hence not moved.}\shan{I am a little lost on the last sentence, especially the sub-sentence after ``even though". Shall we mention our purpose of the ``color'' bits at the beginning of the added part? And maybe add a pointer to `Optimizing for Local Writes.` and omit the detailed explanation here?}\mhr{What about: Additionally, \tool reserves the highest 16 bits in the global address field as ``color'' bits. These bits record the version number of the pointer and play a crucial role in \tool's efficient handling of local writes.}

Next, we discuss how \tool reimplements Rust's ownership operations to realize the distributed coherence protocol.
% \revise{For ease of presentation, this subsection focuses on a simplified version of the protocol. A complete coherence protocol and its proof of memory coherence are available in the appendix.}
For ease of presentation, this subsection focuses on a simplified version of the protocol. A complete coherence protocol and its proof of memory coherence are available in~\cite{distr}.

\begin{algorithm}[t]
\small
\KwIn{A mutable reference $m$ containing a global address $m.g$ and the owner address $m.o$. }
\KwOut{A local memory address to be written to.}
\DontPrintSemicolon
\SetKwFunction{Derefmut}{\textsc{DerefMut}}
\SetKwFunction{Dropmutref}{\textsc{DropMutRef}}
\SetKwFunction{Getcolor}{\textsc{GetColor}}
\SetKwFunction{Clearcolor}{\textsc{ClearColor}}
\SetKwFunction{Appendcolor}{\textsc{AppendColor}}
\SetKwFunction{Islocal}{\textsc{IsLocal}}
\SetKwFunction{Move}{\textsc{Move}}
\SetKwFunction{Copy}{\textsc{Copy}}
\SetKwFunction{Getentry}{\textsc{GetEntry}}
\SetKwFunction{Updateentry}{\textsc{UpdateEntry}}
\SetKwFunction{Insertentry}{\textsc{InsertEntry}}
\SetKwFunction{Write}{\textsc{Write}}

    \SetKwProg{Fn}{Function}{:}{}
    \Fn{\Derefmut{$m$}}{
        \If{$\neg$\Islocal{$m.g$}}{ \label{l:place-check}
            % $c \gets$ \Getcolor{$m.g$}\; \label{l:move-st}
            $m.g \gets$ \Move{\Clearcolor{$m.g$}}\; \label{l:move-mut}
            % $m.g \gets$ \Appendcolor{$m.g$, $c$}\; \label{l:move-ed}
        }
        \Return \Clearcolor{$m.g$}\;
    }

    \Fn{\Dropmutref{$m$}}{ \label{l:dropmutst}
    $c' \gets$ \Getcolor{$m.g$} $+ 1$ \label{l:writeback-mut-st}\;
    \Write{$m.o$, \Appendcolor{$m.g$, $c'$}} \label{l:writeback-mut-ed}\; 
    }\label{l:dropmuted}
\caption{Access logic for mutable references.\label{alg:mut}}
\end{algorithm}

\MyPara{Mutable Borrow.}
Mutable borrow creates a mutable reference that holds exclusive access to the referenced object for writing. Algorithm~\ref{alg:mut} outlines the procedures for both dereferencing and dropping a mutable reference. When performing dereferencing, \tool first checks the object's location (Line~\ref{l:place-check}) and performs direct access if the object's address belongs to the heap partition of the machine A that executes the access. Otherwise, \tool \emph{moves} it to A's heap partition (as opposed to caching it) (Line~\ref{l:move-mut}). The move, conducted in the following three steps, changes the object's global address. \tool (1) copies the object into A's heap at an address $p$, (2) updates the mutable reference with the address $p$, and (3) asynchronously requests the remote server that previously stored the object to deallocate the original object. 

A challenge arises with its original owner \codeIn{Box}, which now becomes a dangling pointer, pointing to an invalid memory location. Fortunately, the integrity of the system is maintained by the single-writer invariant (referenced as Invariant \ref{inv:3}). This invariant ensures that while the mutable reference remains alive, no other entity, including the original owner, can access the data. To ensure correctness, when this new reference is dropped, \tool synchronously updates the original owner \codeIn{Box}, redirecting it to the new address $p$ (Line~\ref{l:writeback-mut-ed}). As a result, the original owner always possesses the latest view of the object. Additionally, all modifications made through this mutable reference are visible in all subsequent accesses, as they necessitate borrowing permission from the updated owner \codeIn{Box}. The single-writer invariant also eliminates the possibility of simultaneous updates to the owner, ensuring that updating the owner is free from concurrency issues.

\begin{algorithm}[t]
\small
\KwIn{A shared immutable reference $r$ containing a global address $r.g$ and a local copy address $r.l$, and a local cache hashmap $\mathit{H}$. }
\KwOut{A local memory address for reading.}
\DontPrintSemicolon
\SetKwFunction{Deref}{\textsc{Deref}}
\SetKwFunction{Dropref}{\textsc{DropRef}}

    \SetKwProg{Fn}{Function}{:}{}
    \Fn{\Deref{$r, H$}}{
        \If{\Islocal{$r.g$}}{ \label{l:islocal-immut}
            \Return \Clearcolor{$r.g$} \label{l:cleartag-immut}\;
        }
        \Else{
            \If{$r.l = \mathit{Null}$}{
                \textsc{\textbf{Atomic}} \{ \\
                    \If{$r.g \in H$}{\label{l:checkcache-immut}
                        $\langle l', cnt \rangle \gets$ \Getentry{$H$, $r.g$}\label{l:cachehit-immut-st}\;
                        $r.l \gets l'$\;
                        \Updateentry{$H$, $r.g$, $\langle l'$, $cnt+1 \rangle$}\label{l:cachehit-immut-ed}\;
                    }
                    \Else{\label{l:cache-exists}
                        $r.l \gets$ \Copy{\Clearcolor{$r.g$}} \label{l:cachemiss-immut-st}\;
                        \Insertentry{$H$, $r.g$, $\langle r.l, 1 \rangle$}\label{l:cachemiss-immut-ed}\;
                    }
                \}\\

                }
            \Return $r.l$\;
        }
    }

    \Fn{\Dropref{$r, H$}}{\label{l:droprefst}
        \If{$r.l \neq \text{Null}$}{
            \textsc{\textbf{Atomic}} \{ \\
            $\langle l'$, $cnt \rangle \gets$ \Getentry{$H$, $r.g$}\;
            \Updateentry{$H$, $r.g$, $\langle l'$, $cnt-1 \rangle$}\label{l:drop-ref-counter}\;
            \}\\
        }
    }\label{l:droprefed}
\caption{Access logic for immutable reference.\label{alg:imm}}
\end{algorithm}

\MyPara{Immutable Borrow.}
Immutable borrowing allows concurrent reads to the same object from immutable references on the same or different servers. As detailed in Algorithm~\ref{alg:imm}, \tool handles the dereferencing of immutable references by first checking the object's location (Line~\ref{l:islocal-immut}). For remote objects, \tool creates a local copy in the \emph{per-node read-only ``cache''} and records its local address in the reference's extension field (see Figure~\ref{fig:pointer-layout}). This preserves the original global address of the object, ensuring that any new immutable reference\textemdash whether it is derived from the owner \codeIn{Box} or from another immutable reference\textemdash can always access the original object from the global heap.

As opposed to being a separate memory space, our ``cache'' provides a ``virtual'' aggregation of all local copies maintained on each server. These copies reside in the regular heap, managed by a per-node hashmap $H$. This hashmap maps each global address to a pair of its local address and the number of local immutable references to the local copy. To prevent redundant copies of an object on the same server, \tool checks the hashmap $H$ before creating a new local copy (Line~\ref{l:checkcache-immut}). If a local copy is already present, \tool increments its reference count in $H$ and updates the extension field in the immutable reference to point to this copy (Lines~\ref{l:cachehit-immut-st}--\ref{l:cachehit-immut-ed}). If no existing copy is found, a new one is created (Lines~\ref{l:cachemiss-immut-st}--\ref{l:cachemiss-immut-ed}). Since the hashmap uses objects' global addresses as keys, if an object has been modified by another server since its last read, its global address must have changed, making cache lookup fail even if a (stale) local copy exists.

\tool actively updates the reference count of each local copy when an immutable reference is either dereferenced or dropped, as outlined in Lines~\ref{l:cachehit-immut-ed} and~\ref{l:drop-ref-counter}. Utilizing these counts, the \tool runtime periodically scans the ``cache'' and lazily reclaims unreferenced copies (\ie, those with a zero reference count) under memory pressure (\S\ref{sec:rtsys:apprt:allocator}). This mechanism, in conjunction with the safe borrowing invariant (\ref{inv:2}), prevents the local cache from memory leaks or illegal accesses.

\MyPara{Owner Access without Borrow.} \tool treats a direct memory access via the owner \codeIn{Box} as a pair of mutable/immutable borrow and return. Depending on the reference type,
\tool uses the extension field of \codeIn{Box} accordingly and executes the read/write dereferencing logic. A special case arises when a mutable owner is immutably borrowed and becomes immutable until all borrowed references return. In this case, the owner can only cache the object during the borrow and delay the move until the borrow finishes. This would not create any correctness issues because the owner cannot be used for write access during this period.

\MyPara{Ownership Transfer.} Similar to Rust, \tool does not move the actual value during the transfer and only copies the \codeIn{Box} pointer. \tool additionally checks and resets the pointer's extension field and frees the cached copy in the executing machine's cache to avoid cache leakage.  %\hx{local copy field?}\yq{fixed.}

\MyPara{Memory Deallocation.} Like Rust, \tool tracks the lifetime of an object via its owner. Given that ownership transfer is implemented by only evicting the cached copy of the object (without changing its global presence), the memory safety of \tool's global heap is preserved by the singular owner invariant (\ref{inv:1}). In other words, \tool still guarantees that when an object's owner goes out of scope, the object must be unreachable (and dead) and can be safely deallocated.

\MyPara{Consistency Model.\label{sec:pa:consistency-model}}
Our protocol, together with Rust's ownership model, offers \emph{sequential consistency} for cross-server memory accesses in safe Rust programs (\ie, following the original Rust, no guarantees can be provided when Rust \codeIn{Unsafe} is used), which is a strong consistency order. Therefore, it allows any safe Rust program to preserve its memory consistency on DSM. Sequential consistency necessitates a coherent memory system, requiring not only the SWMR invariant but also the \emph{data-value} invariant~\cite{prime-coherence-book2020}. In simple terms, the data-value invariant requires that the latest write to a value is immediately visible to subsequent readers. As discussed earlier, \tool's protocol moves an object upon a write and updates the owner immediately. Therefore, the latest value is globally visible after each mutable borrow finishes. Subsequent read accesses, either in the Owned state or the Shared state, are hence guaranteed to see the moved object and read its latest value. 

\begin{algorithm}[t]
\small
\DontPrintSemicolon
\SetKwProg{Fn}{Function}{:}{}
\Fn{\Getcolor{$g$}}{
    \Return $g \gg 48$\;
}

\Fn{\Clearcolor{$g$}}{
    \Return $g$ \texttt{\&} $((1 \ll 48) - 1)$\;
}

\Fn{\Appendcolor{$g$, $c$}}{
    \Return \Clearcolor{$g$} \texttt{|} ($c \ll 48$)\;
}
\caption{Utility functions for pointer coloring.\label{alg:util}}
\end{algorithm}

\MyPara{Optimizing for Local Writes.}
A special case is that a server issues a write to an object that resides in its own heap partition. While the coherence protocol still guarantees safety, requiring moving an object in its local heap each time it is written clearly brings inefficiencies. 
To optimize for local writes, \tool adopts a pointer-coloring method, inspired by the design of concurrent garbage collectors in a managed runtime system such as JVM~\cite{zgc,mako-pldi22}. Several utility pointer coloring functions are shown in Algorithm~\ref{alg:util} which are used when dereferencing and dropping a reference. We reserve the first 16 bits of a global address as a ``color''. The color value stored in the object's owner gets incremented upon the expiration of a mutable reference, as detailed in Lines~\ref{l:writeback-mut-st}--\ref{l:writeback-mut-ed} in Algorithm~\ref{alg:mut}. Any subsequent immutable borrow would look up the cache with the object's global address. Even if the actual address remains the same, its color changes if a write has occurred. As such, the lookup would not return any stale copy from the local cache.

The 16-bit \codeIn{color} field may overflow when the pointer keeps being borrowed for local writes on the same server. \tool implements a \textit{move-on-overflow} strategy that moves the object to a new address and resets its color to zero once the maximum color value is reached ($2^{16}$), thereby preventing overflow and maintaining system integrity and performance.

% A special case is that a server issues a write to an object that resides in its own heap partition. While the coherence protocol still guarantees safety, requiring moving an object in its local heap each time it is written clearly brings inefficiencies. 
% To optimize for local writes, we devise an optimization that does not move an object if writes are local, leaving the object's global address unchanged. However, this may direct subsequent immutable references to this object from other servers to use stale copies of the object in their local caches.  \tool solves this problem using a pointer-coloring method, inspired by the design of many concurrent garbage collectors~\cite{zgc,mako-pldi22}. We reserve the first 16 bits of a global address as a ``color''. The color value stored in the object's owner gets incremented upon the expiration of a mutable reference. Any subsequent immutable borrow would still look up the cache with the object's global address. However, even if the actual address remains the same, its color changes if a write has occurred. As such, the lookup would not return any stale copy from the local cache.

%using the updated tag and global address, effectively avoiding the retrieval of obsolete data. This mechanism offers a much more efficient solution than the copy-on-write approach.

\MyPara{Writing Unsafe Code in \tool.\label{sec:discussion:unsafe}}
Rust allows developers to bypass compiler safety checks and write \emph{unsafe} code for low-level operations such as accessing raw pointers and mutating shared variables at their own risk~\cite{jung2017rustbelt, qin2020understanding}.
Since \tool relies on SWMR semantics enforced by Rust's ownership types, \tool ensures consistency and memory safety only in the ``safe'' Rust code. \tool does not cache objects in unsafe code but allows developers to implement their own cache. Developers must ensure that they do not violate consistency in unsafe code blocks where type safety is not enforced. This caution mirrors practices in other managed languages, like native code in Java and unsafe code in C\#. To assist developers, \tool offers primitives such as \codeIn{dalloc}, \codeIn{dread}, and \codeIn{dwrite} for managing data on the global DSM heap.

\subsubsection{Adapting Rust Standard Libraries \label{sec:design:pa:std-lib-support}}
To further reduce the barrier for programs to run distributively, we reimplement several standard Rust libraries atop \tool's core memory constructs covering four categories: threading for distributed computation (\codeIn{std::thread}), inter-thread channel for communication (\codeIn{std::sync::mpsc}), reference-counted pointers for ownership sharing (\codeIn{std::sync::Rc} and \codeIn{std::sync::Arc}), and shared-state locks for concurrency control (\codeIn{std::sync::Mutex} and \codeIn{std::sync::atomic}).
% The Rust Standard Library is pivotal in Rust programming, offering a broad spectrum of essential functionalities. When transitioning to a distributed shared memory (DSM) environment, several of these components necessitate reconstruction. \tool's reengineered standard library together its redesigned core memory management constructs, can seamlessly scale a single-machine Rust program to distributed settings with great efficiency. This subsection will delve into these adaptations.

\MyPara{Threading.}
\tool's threading library enables Rust threads to run distributively with two major adaptations. First, it enables distributed thread launching by re-implementing the \codeIn{spawn} interface. Internally, it captures the thread body as a closure during compile time and forwards it to the runtime. During execution, the runtime launches the thread according to each server's load (details in \S\ref{sec:rtsys:apprt:thd-scheduler}). Second, \tool performs implicit ownership transfers between the parent and the child threads at the start or the end of the child thread execution. Thanks to the distributed ownership transfer support provided by \tool's memory model, the implementation in the threading library is hidden from developers and preserves type soundness and memory safety. Additionally, \tool is compatible with advanced thread utilities such as \codeIn{thread::scope}, which allows for the spawning of scoped threads that can borrow non-static data. These utilities ensure that all threads are joined at the end of their scope and can internally utilize \tool's functions for spawning and joining threads, thus extending their applicability to the distributed setting.

\MyPara{Inter-Thread Channel.} \tool extends Rust's channel to connect two distributed threads for message passing. \tool internally builds a network-based message queue for cross-server messages. Benefiting from the shared global heap, \codeIn{Box} pointers and references can be safely copied and remain valid across servers. Therefore, the sender can push an object into the channel \emph{as is} without serialization, even if it may contain \codeIn{Box} pointers. 
\tool forwards the object binary bytes to the receiver over the network, and the receiver can recover the object from the binary by direct type conversion without deserialization.

\MyPara{Ownership Sharing.}
Rust allows multiple owners to share an object via reference-counted smart pointers, which count the number of live owners. In this case, smart pointers only have read access, and the object lifetime terminates when all owners die and the reference count hits zero. \tool does not require special treatment for \codeIn{Rc} as it only allows ownership sharing inside a single thread. For \codeIn{Arc} which shares ownership among multiple threads, \tool handles it in a similar way to immutable references with on-demand local caching and lazy eviction.
%\yq{updated eviction policy to lazy.}

% Rust's standard library offers the Rc<T> and Arc<T> smart pointers for enabling multiple ownership of a value through reference counting. These pointers track the number of active references to ensure safe deallocation when no references remain. For single-threaded scenarios, the Rc<T> smart pointer remains unchanged, as it is inherently unsuitable for multi-threaded or distributed environments. In contrast, the Arc<T> pointer, designed for multi-threaded contexts, extends its utility to DSM settings. In these scenarios, multiple owners across various threads and servers can coexist, but they are limited to read-only access. This restriction aligns the behavior of Arc<T> closely with immutable references, with a key difference in \tool's handling of data access through them: when the reference count of a value drops to zero, the system does not immediately deallocate its cache copy. Instead, it retains this copy locally to optimize potential future reads, given the read-only nature of the data. \yiying{can there be any data leak (security concern) here?}
% This strategy enhances efficiency, ensuring that cached data remains readily available unless memory pressure necessitates its release. In situations where memory usage on a server exceeds 90\%, the system employs the LRU policy to selectively purge cached copies with a reference count of zero. This approach balances memory management and data access efficiency in DSM environments, leveraging the unique characteristics of Rust's smart pointers.

\MyPara{Shared-State Concurrency.}
Rust supports shared-state concurrency, primarily through its atomics and mutexes, where threads commonly share an atomic-typed value or one mutex via ownership sharing (\ie, \codeIn{Arc}). Unfortunately, the ownership model cannot type check concurrent read/write to shared states. Hence, Rust relies on an \emph{unsafe} implementation in its standard library. \S\ref{sec:design:pa:memory-management} already provides a general discussion on writing unsafe code in \tool, and here we focus on \tool's implementation for distributed shared states.

Shared states create a unique challenge for \tool, as they may be replicated on multiple servers and those states must be synchronized among these servers. For example, an \codeIn{Arc<AtomicBool>} may be replicated across different servers and used independently, causing multi-version issues if not synchronized properly. \tool addresses this inconsistency by allocating the actual value on the global heap and storing only the \codeIn{Box} pointer in \codeIn{atomic} types. This design allows atomics to be freely moved or replicated across servers while keeping a single version of the actual value. To synchronize concurrent operations on atomics, \tool adapts methods of atomic types to forward the operation as a message to the server storing the actual value, which serializes all operations with unsafe logic similar to the original Rust to guarantee atomicity and memory consistency. 
%\qd{Does our argument that guarantees sequential consistency hold here? Are we missing other complexities with message ordering / memory barriers here? }\yq{How about now? We ensure the same consistency guarantee as the original Rust by serializing all operations. I feel it may be a bit complicated to discuss the consistency of atomic operations here.}\qd{Better now. Be ready for reviewer comments on this though. :-)}
Similarly, \tool implements \codeIn{Mutex} by allocating its metadata and owned object on the global heap and leaving only \codeIn{Box} pointers in the mutex struct. Concurrent operations on mutexes are serialized on the server storing the mutex.

\subsubsection{Affinity Annotations\label{sec:design:pa:affinity-annotations}}

To further improve performance, \tool allows developers to provide optional data affinity semantics via annotations. These annotations are useful for many datacenter applications that make extensive use of object-oriented data structures that require \emph{pointer-chasing} to access. For instance, Memcached~\cite{memcached} uses a chained hash table where each hash bucket stores its KV pairs with a linked list. To find one KV pair from a bucket, Memcached has to iterate the linked list following the node pointers. However, frequent pointer chasing is unfavorable in a distributed setting, where each pointer dereference incurs additional runtime checks and potential cross-server traffic. It would be beneficial for the runtime to colocate them on the same server and schedule the computation there.

\begin{lstlisting}[language=Rust, float=t, floatplacement=t, 
caption={A linked list implementation with \codeIn{TBox} in \tool. The use of \codeIn{TBox} ties list nodes one by one. Iterating a list will fetch all nodes together (if they are on another server), and henceforth accessing any node is guaranteed local.},
belowskip=-2em,
label={lst:tbox-linked-list}]
pub struct Node { val: i32, next: Option<TBox<Node>>, } /*@\label{lst:list:tbox-dec1}*/
pub struct List { pub head: Option<Box<Node>>, } /*@ \label{lst:list:tbox-dec2} */
impl List { 
  pub fn sum(&self) -> i32 { /*@ \label{lst:list:sum-stt} */
    let mut total: i32 = 0;
    if let Some(r) = &self.head {
      let mut node = &**r; // Fetch whole list to local./*@\label{lst:list:fetch-list}*/
      loop { // Iterate every list node. /*@ \label{lst:list:loop-stt} */
        // Accessing node is guaranteed local.
        total += (*node).val; 
        if let Some(next) = &node.next {
          node = &**next;
        } else { break; }
      } /*@ \label{lst:list:loop-end} */
    }
    total
  } /*@ \label{lst:list:sum-end}*/
}

\end{lstlisting}

\MyPara{Data-Affinity Pointer.}
To expose data affinity for clustered placement, \tool includes a new pointer type \codeIn{TBox} for developers to ``tie'' a heap object to its owner. \codeIn{TBox} shares the same interfaces as the ordinary \codeIn{Box} and can be used as a drop-in replacement for \codeIn{Box}. However, \codeIn{TBox} enforces that the pointed-to object must reside on the same server as its owner. In other words, when its owner object is copied or moved, the object referenced by \codeIn{TBox} will be copied or moved as well. 
\codeIn{TBox} can be used in a nested manner to allow a large union of objects to be co-located. 
% The singular owner invariant ensures the existence of the root object for nested \codeIn{TBox}es.
%Developers can use \codeIn{TBox} to group objects that are closely correlated, and 
The \tool runtime fetches (\ie, copies or moves) them together in a single batch, leading to fewer network round-trips and higher throughput.
\codeIn{TBox} can also be assigned to and owned by a stack variable, in which case the referenced object is pinned onto the heap partition of the server that hosts the stack and cannot be moved. Dereferencing a \codeIn{TBox} is thus guaranteed to be a local access\textemdash \tool optimizes it by skipping the runtime check.

Listing~\ref{lst:tbox-linked-list} presents a linked list implementation using \codeIn{TBox}. 
%Each list node holds an integer value and a pointer to the next node, and the %list head points to the first node. 
Our linked list uses \codeIn{TBox} (Line~\ref{lst:list:tbox-dec1}) to specify the data affinity between consecutive nodes.
% , and between the list struct and its head node.
As a result, all list nodes are chained with \codeIn{TBox}, forming an affinity group. Line~\ref{lst:list:sum-stt}--\ref{lst:list:sum-end} define a \codeIn{sum} function that calculates the total sum of all node values. Assuming the list is non-empty, Line \ref{lst:list:fetch-list} dereferences the pointer to the head node,
% pointer to a stack value, 
and the \tool runtime checks the location of the head node and fetches the entire list of nodes together if they are not local. Next, accessing each node inside the loop body (Line \ref{lst:list:loop-stt}--\ref{lst:list:loop-end}) is guaranteed local and hence skips runtime checks. Compared to using \codeIn{Box} directly, \codeIn{TBox} makes both data fetching and accessing more efficient.

\begin{lstlisting}[language=Rust, float=t, floatplacement=t, 
caption=A distributed accumulator can leverage \codeIn{spawn\_to} to offload a thread to the server where \codeIn{a.val} locates.,
belowskip=-2em,
label={lst:spawnto-accumulator}]
fn main() {
  let val: Box<i32> = Box::new(5);
  let mut a = Accumulator{val};
  let remote_add = spawn_to(a.val, move || 
    a.add(10)).join(); // a.val == 15 /*@ \label{lst:spawnto:line:remote-add} */
}
\end{lstlisting}

\MyPara{Data-Affinity Thread.}
To expose the affinity between data and computation for thread scheduling, \tool extends its threading library with a \codeIn{spawn\_to} interface. \codeIn{spawn\_to} mirrors the ordinary \codeIn{spawn} interface to spawn a new thread but takes an additional \codeIn{Box} pointer argument, which indicates where the thread should be created. The runtime checks where the \codeIn{Box} points to and creates the new thread on that same server. A common practice to use \codeIn{spawn\_to} is to pass the mostly-accessed object as the location indicator.
Listing~\ref{lst:spawnto-accumulator} presents how the distributed accumulator (shown in Listing~\ref{lst:dist-accumulator-sample}) can use \codeIn{spawn\_to} to offload a thread to the same server as \codeIn{a.val} resides. Line~\ref{lst:spawnto:line:remote-add} hence performs local dereference to \codeIn{a.val} inside \codeIn{a.add()}.

% \begin{lstlisting}[language=Rust, style=boxed, caption=Code example of thread spawning interface, label={lst:thread} ]
% fn main() {
%     let d: Box<D> = thread::spawn(move || {Box::new(D {..})}).join();
%     let h = thread::spawn_to(d, move || {compute_func(d)});
%     h.join();
% }
% \end{lstlisting}

% \mhr{To update with a better example}

% Beyond the data affinity, the relationship between compute operations and data location is equally vital. \tool's thread manager, as outlined in section~\ref{sec:std}, mirrors Rust's native thread interface for ease of use. \tool's default thread distribution method, based solely on server resource availability, still has room for performance optimization, particularly when data and compute operations are not co-located. 
% To address this, we introduce a new interface that allows programmers to hint at the primary data dependencies of a thread. This interface, illustrated in listing~\ref{lst:thread}, accepts an additional argument (e.g., $Box\langle T\rangle>$), indicating the data structure most accessed by the new thread.

% This approach allows for more informed thread scheduling. If the server hosting the indicated data has sufficient compute resources, the thread is preferentially scheduled there, minimizing data transfers. If not, \tool reverts to its default load-balancing approach. This nuanced scheduling mechanism can enhance performance by aligning compute operations with their primary data sources.

\subsection{\tool Runtime System\label{sec:design:runtime-system}}

\tool's runtime system is the core component that manages memory and compute resources. It includes 
a runtime library (\S\ref{sec:rtsys:apprt}) that is linked to each application and launched on each server and a cluster-wise global controller (\S\ref{sec:rtsys:controller}). 

\subsubsection{Application-Integrated Runtime}
\label{sec:rtsys:apprt}
The runtime library consists of a communication layer 
% (\S\ref{sec:rtsys:apprt:comm-layer}) 
to support inter-server coordination and data transfer, a heap allocator 
% (\S\ref{sec:rtsys:apprt:allocator}) 
to manage the heap partition and the read-only cache, and a thread scheduler 
% (\S\ref{sec:rtsys:apprt:thd-scheduler}) 
to launch and schedule application threads. 

\MyPara{Communication Layer.}
\label{sec:rtsys:apprt:comm-layer}
The \tool runtime uses its communication layer to support the cache coherence protocol and cross-server memory accesses.
The communication layer consists of a control plane and a data plane, both built with RDMA. The control plane leverages \emph{two-sided verbs} to send and receive small control messages, and the receiver can perform the coherence logic upon receiving the message to minimize the end-to-end latency. The data plane, in contrast, is specialized for bulky data transfer with one-sided verbs. It fetches an object as a whole with a single RDMA message upon pointer dereferencing without interrupting the target server, minimizing both latency and CPU usage.
%\yj{What's the rationale behind using different verbs for control/data plane. Maybe provide a couple of citations?}

\MyPara{Heap Allocator.}
\label{sec:rtsys:apprt:allocator}
The \tool runtime provides standard memory allocation interfaces and always returns global addresses to the upper-level language abstractions. It prioritizes local memory allocation as long as the local heap partition has sufficient space. This strategy improves data locality by colocating an object with its creating thread. 

When the local heap partition is depleted, \tool queries the global controller and allocates memory on the most vacant server.
For remote memory allocation, it forwards the request to the target server by sending a message through the communication layer and returns the allocated address to the user. 
Memory deallocation follows a similar logic but it bypasses the controller and finds the server directly via the object's global address.
The allocator does not reserve separate space for the local cache. Instead, it manages the cache as part of the local heap partition and always allocates cached entries in the local heap partition. Under memory pressure, the allocator will scan the local cache and evict entries that are no longer used (\ie, reference count hits zero).

\MyPara{Thread Scheduler.}
\label{sec:rtsys:apprt:thd-scheduler}
The \tool thread scheduler runs in the user space and schedules threads locally to maximize CPU utilization. It also provides thread migration functionalities, facilitating the global controller to balance load between busy and vacant servers. 

The scheduler represents a newly created user thread as a closure, which includes a function pointer and a set of initial arguments (\ie, references). It collaborates with the global controller to allocate a unique stack space for a thread (see Figure~\ref{fig:address-space}), and starts the thread by executing the closure.

The scheduler adopts the method of cooperative multitasking and context switches between threads \emph{non-preemptively}. A running thread yields its control flow proactively when developers call \codeIn{await} or reactively upon long-latency operations. 
Similarly to other systems~\cite{nelson2015latency, shenango-nsdi19, rust-tokio}, our scheduler handles context switches as function calls, which allow \tool to save only a few registers per thread.

The scheduler supports creating/migrating a thread to another server as well. To migrate a thread, \tool sends its function pointer, the saved register state, and its stack to the target server. Because each thread reserves its stack address range globally, \tool can copy the stack across servers without changing its address. Therefore, the thread can be easily resumed by directly calling the function pointer with the saved register state on the target server. \tool generates code for state transmission during the compile time for the scheduler to call upon thread migration.

\subsubsection{Global Controller}
\label{sec:rtsys:controller}

The controller runs as a daemon process on the machine where the program is launched. It manages cluster resources and coordinates memory allocation and thread migration. It periodically pings each server to probe and record its resource usage (CPU and memory). It controls resource allocation in cooperation with the \tool runtime on each server. When allocating memory or creating a thread, the runtime will first query the controller, which chooses a target server following adaptive policies (discussed later), and then coordinate with the runtime on the target server to perform the actual operation. The controller also maintains a global table to track the location of each thread; the table is queried and updated by the scheduler when migrating a thread.

During program execution, servers may run into imbalanced loads when objects get relocated or new threads are created. \tool balances the load of each server by \emph{migrating} threads from the busy server to less occupied ones, following an adaptive policy to minimize cross-server memory accesses. If a server is about to run out of memory (>90\% memory usage), the controller keeps migrating the thread that consumes the most local heap memory until the pressure is resolved. If the server is under compute congestion (>90\% CPU usage), the controller migrates threads that frequently access remote objects. The thread is then moved to the server it accesses the most unless the target server is also overloaded, in which case it moves to a vacant server instead.
\subsubsection{Fault Tolerance
\label{sec:discussion:fault}} 
In \tool, the global heap can be replicated to tolerate failures. Replication creates copies for each heap partition at the same virtual address on backup servers. Threads, in contrast, are not replicated for efficiency and are only executed with the primary global heap. To maintain a synchronized view between the primary heap partition and its backup copy, a thread must additionally write back to the backup partition after each mutable borrow. However, our insight is that the thread can batch modifications to an object and delay the write-back until the object ownership is transferred to another server, which is the time point that the object becomes visible to threads on other servers. When a server with a primary heap partition fails, the controller will automatically promote its backup server to the primary and add a new backup server.
\section{Implementation}
\label{sec:impl}

% \mhr{\sout{The majority of \tool was implemented in Rust except for its communication library which is in C. Our implementation includes 5.6K SLOC of Rust code for the programming abstraction (\S\ref{sec:design:abstraction}), 3.5K SLOC for the runtime (\S\ref{sec:rtsys:apprt}), and 2.3K SLOC for the global controller (\S\ref{sec:rtsys:controller}). }}

The majority of \tool was implemented in Rust except for its communication library which is in C. We implemented \tool's core language constructs as three Rust types (\ie, \codeIn{struct}): \codeIn{Ref<T>}, \codeIn{MutRef<T>}, and \codeIn{DBox<T>}. They serve as the counterpart for the original Rust \codeIn{\&T}, \codeIn{\&mut T}, and \codeIn{Box<T>}, respectively. We implemented the coherence protocol with traits on these types, including \codeIn{Copy}, \codeIn{Clone}, and \codeIn{Drop}, which are automatically embedded into the program source code and executed when references/pointers are created or destroyed. To support unmodified Rust programs, we changed the Rust compiler and added additional compilation passes to transform Rust references and \codeIn{Box} pointers into corresponding types in \tool.

Our communication layer links \codeIn{libibverbs} directly for fast and kernel-bypassing RDMA networking. We implemented a low-level C library that covers basic connection establishment and exposes high-level Rust interfaces for various RDMA verbs, including \codeIn{RDMA\_READ}, \codeIn{RDMA\_WRITE}, \codeIn{RDMA\_SEND}, \codeIn{RDMA\_RECV}, \codeIn{ATOMIC\_FETCH\_AND\_ADD}, and \codeIn{ATOMIC\_CMP\_AND\_SWP}.
We primarily utilize one-sided \codeIn{READ} and \codeIn{WRITE} verbs for data transfers between servers, as they outperform the two-sided \codeIn{SEND}/\codeIn{RECV} counterparts\textemdash one-sided operations bypass the CPU and OS at the receiver side, whereas two-sided operations require the receiver to pre-post \codeIn{RECV} verbs and await notification upon message arrival. 
For instance, when a remote object is accessed via mutable references, \tool copies the object to local memory using the \codeIn{READ} verb. Upon dropping the reference, \tool updates the original owner \codeIn{Box} to reflect the new address, a process executed using the \codeIn{WRITE} verb. Conversely, two-sided \codeIn{SEND}/\codeIn{RECV} verbs are utilized for control message exchanges, such as establishing connections across servers. Atomic verbs \codeIn{ATOMIC\_FETCH\_AND\_ADD}, and \codeIn{ATOMIC\_CMP\_AND\_SWP} are primarily utilized for implementing shared states (e.g., atomic types and mutexes).
\tool uses the RC (reliable connection) transport type to ensure reliable transmission and strict message ordering.

Our heap allocator implementation piggybacks Rust's original allocator and aligns its virtual address range with the heap partition range. Our thread scheduler was built upon Tokio~\cite{rust-tokio} for its efficient user thread and cooperative scheduling integration. 
% We enforce a fixed size (default 2MB) and 16-byte alignment for thread stacks to be compatible with Linux. 
The global controller is responsible for managing all threads in the cluster and padding their stacks to avoid address overlapping.

\section{Limitations}
% \revise{\tool's design has three limitations. First, although \tool permits the use of unsafe code, its consistency guarantees are only applicable to safe Rust code. In unsafe code blocks, developers are responsible for ensuring consistency themselves. Second, to facilitate the implementation of the distributed global heap and distributed thread spawning, we disable address space layout randomization (ASLR). Future modifications to the OS kernel could allow \tool applications to have the same randomized address layout across all machines. Finally, \tool's superior performance relies on SWMR semantics exposed by applications. In cases where data is mostly under shared states (such as \codeIn{Mutex}), \tool degenerates into a traditional DSM system; all concurrent accesses to the same data have to be centralized and serialized by the server responsible for the shared states. However, such scenarios contradict Rust's recommended programming practices. 
% }
\tool's design has three limitations. First, although \tool permits the use of unsafe code, its consistency guarantees are only applicable to safe Rust code. In unsafe code blocks, developers are responsible for ensuring consistency themselves. 
% Second, to facilitate the implementation of the distributed global heap and distributed thread spawning, we have temporarily disabled address space layout randomization (ASLR). In future versions, \tool plans to centralize the randomization of stack and heap address allocation through its global controller. This will ensure a consistent view of the randomized address layout across all nodes.
% Consequently, ASLR can be re-enabled on each machine for enhanced security.
Second, \tool's superior performance relies on SWMR semantics exposed by applications. In cases where data is mostly under shared states (such as \codeIn{Mutex}), \tool degenerates into a traditional DSM system; all concurrent accesses to the same data have to be centralized and serialized by the server responsible for the shared states. However, such scenarios contradict Rust's recommended programming practices. 
Finally, the current implementation of \tool does not support address space layout randomization (ASLR) yet, and we have temporarily disabled it. However, \tool's design is compatible with ASLR as long as \tool threads share the same randomized address space layout on each server. This can be achieved by delegating the randomization of stack and heap address allocation in \tool to its global controller, a feature that will be supported in future versions of \tool.

\section{Evaluation\label{sec:eval}}
%Our evaluation aims to answer the following questions:
%\begin{enumerate}[nosep, left=0pt, itemsep=0pt, topsep=0pt]
%\item How well does \tool perform compared to state-of-the-art DSMs and how well does it scale? (\S\ref{sec:eval:overall-perf})
%\item Can \tool's affinity annotations effectively expose data affinity semantics and improve application performance? (\S\ref{sec:eval:drilldown})
%\item What's the performance impact of \tool's runtime choices and policies? (\S\ref{sec:eval:rt-check}--\S\ref{sec:eval:drilldown})
%\end{enumerate}

\MyPara{Setup.}
We evaluated our system on an 8-node cluster, where each node was equipped with dual Intel Xeon E5-2640 v3 processors (16 cores), 128GB of RAM, and a 40 Gbps Mellanox ConnectX-3 InfiniBand network adapter, connected by a Mellanox 100 Gbps InfiniBand switch. All servers ran Ubuntu 18.04 with kernel 5.14. We disabled hyperthreading, CPU frequency scaling, OS security mitigations
% and ASLR 
in accordance with common practices~\cite{qiao2023hermit, ruan2023nu}.

\MyPara{Methodology.}
We compared \tool with two state-of-the-art DSM systems, GAM~\cite{cai2018efficient} and Grappa~\cite{nelson2015latency}.
For a fair comparison, we ported the evaluated applications to each baseline system and invested extensive effort in tuning parameters to achieve their best possible performance. GAM offers ordinary object read/write interfaces, and we exported it as a library to Rust and hooked pointer dereferencing to use GAM's API without program modification. Grappa, in contrast, offers a drastically different programming abstraction that requires rewriting the program to access shared memory via \emph{delegation}. Therefore, we re-implemented applications in C++ and re-structured them using Grappa's abstractions.
%for their optimal performance.

% In our evaluation, we run \tool on four applications against two state-of-the-art software-based Distributed Shared Memory (DSM) systems: Grappa and GAM.\yj{[Citation?]}
% Grappa is designed to operate without a cache. Its core philosophy is to delegate all computational tasks, including data access, directly to the server storing the data. This approach contrasts with the traditional method of transferring data to the computational site. Grappa leverages user-mode MPI for message passing, which internally utilizes RDMA verbs.
% Conversely, GAM implements a directory-based cache coherence protocol built on top of RDMA. It uses two dedicated background threads for managing cache coherence events in addition to the standard application threads. For memory management, GAM offers a suite of primitive interfaces, including \codeIn{malloc}, \codeIn{free}, \codeIn{read}, \codeIn{write}, and \codeIn{mfence}, allowing programmers to handle memory operations.

\subsection{Applications} 
We evaluated four representative datacenter applications covering a wide range of use cases and resource demands, including data analytics, microservices, scientific computation, and key-value storage, as shown in Table~\ref{tab:eval-apps}. 

% \begin{table}[t]
% \centering
% \begin{adjustbox}{max width=\linewidth}
% \small
% \begin{tabular}{ccccr}
% \toprule
% \thead{Application} & \thead{Dataset}  & \thead{Memory \\ (GB)} & \thead{R/W \\ Ratio} & \thead{Comp. Intensity \\ (cycles/byte)} \\ \midrule
%  DataFrame~\cite{polars}  & h2oai~\cite{h2oaibenchmark}  & 64 & 72.5\% & 110.13 \\ \midrule
 
%  SocialNet~\cite{gan2019deathstarbench} & Socfb-Penn94~\cite{penn94} & 64 & 77.8\% & 86.09 \\ \midrule
%  GEMM~\cite{blackford2002blas} & LAPACK\cite{lapack-bench} & 96 & 69.8\% & 300.63 \\ \midrule
%  KV Store~\cite{cai2018efficient} & YCSB~\cite{cooper2010benchmarking} & 48 & 58.7\% & 48.15\\
% \bottomrule
% \end{tabular}
% \end{adjustbox}
% % \vspace{-0.8em}
% \caption{\label{tab:eval-apps} Applications used in the evaluation.
% }
% \vspace{-1em}
% \end{table}
\begin{table}[t]
\centering
\begin{adjustbox}{max width=\linewidth}
\small
\begin{tabular}{ccccr}
\toprule
\thead{Application} & \thead{Dataset}  & \thead{Memory \\ (GB)} & \thead{Comp. Intensity \\ (cycles/byte)} \\ \midrule
 DataFrame~\cite{polars}  & h2oai~\cite{h2oaibenchmark}  & 64 & 110.13 \\ \midrule
 
 SocialNet~\cite{gan2019deathstarbench} & Socfb-Penn94~\cite{penn94} & 64 & 86.09 \\ \midrule
 GEMM~\cite{blackford2002blas} & LAPACK\cite{lapack-bench} & 96 & 300.63 \\ \midrule
 KV Store~\cite{cai2018efficient} & YCSB~\cite{cooper2010benchmarking} & 48 & 48.15\\
\bottomrule
\end{tabular}
\end{adjustbox}
% \vspace{-0.8em}
\caption{\label{tab:eval-apps} Applications used in the evaluation.
}
\vspace{-1.5em}
\end{table}

\begin{figure*}[t]
    \centering
    \small
    
    \begin{tabular}{cccc}
         \includegraphics[width=0.24\linewidth]{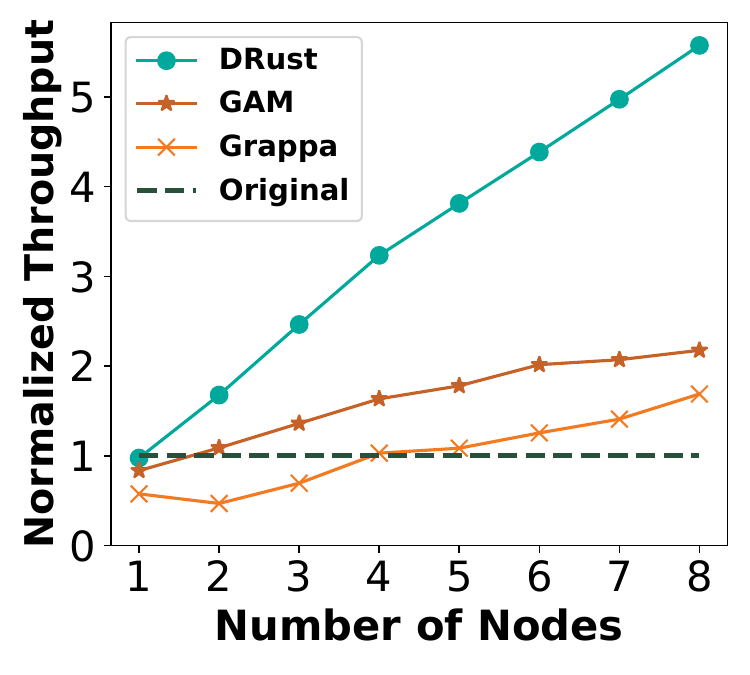}&  \includegraphics[width=0.24\linewidth]{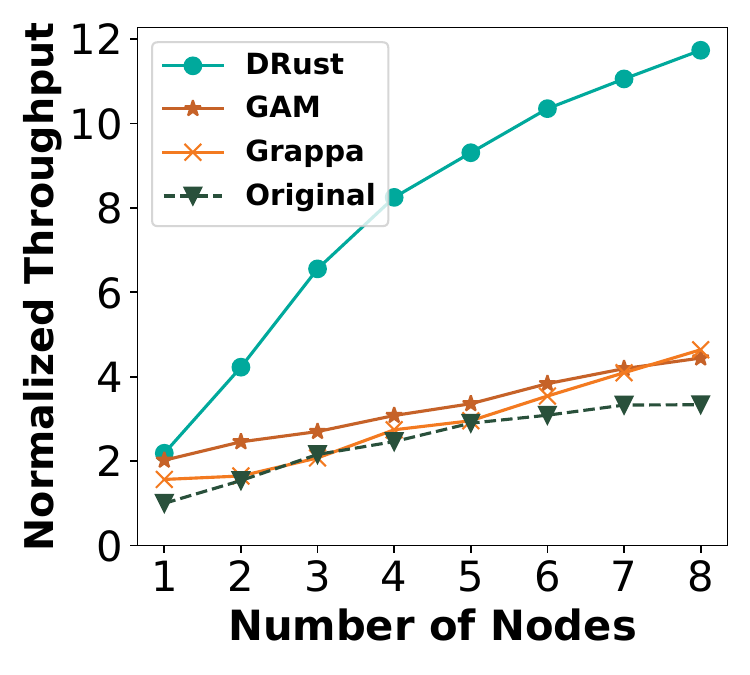}&  \includegraphics[width=0.24\linewidth]{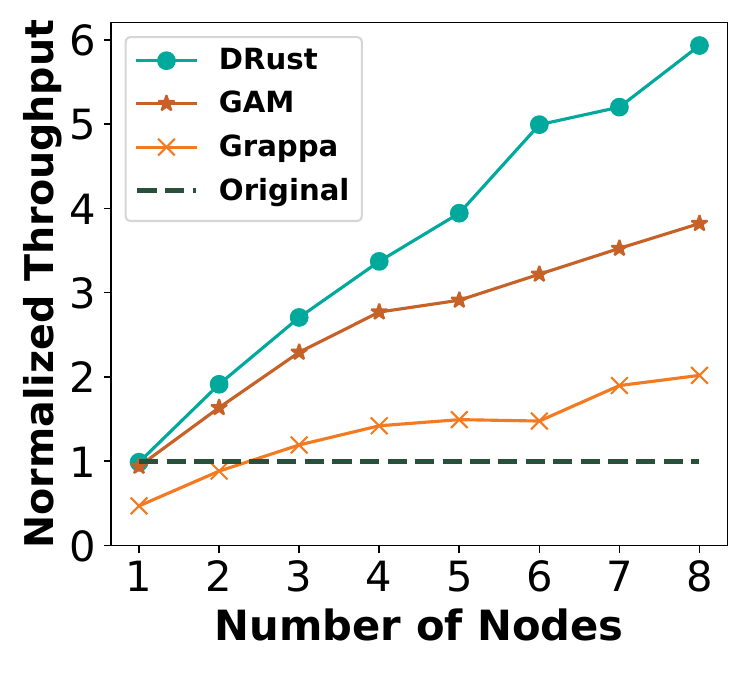} & 
         \includegraphics[width=0.24\linewidth]{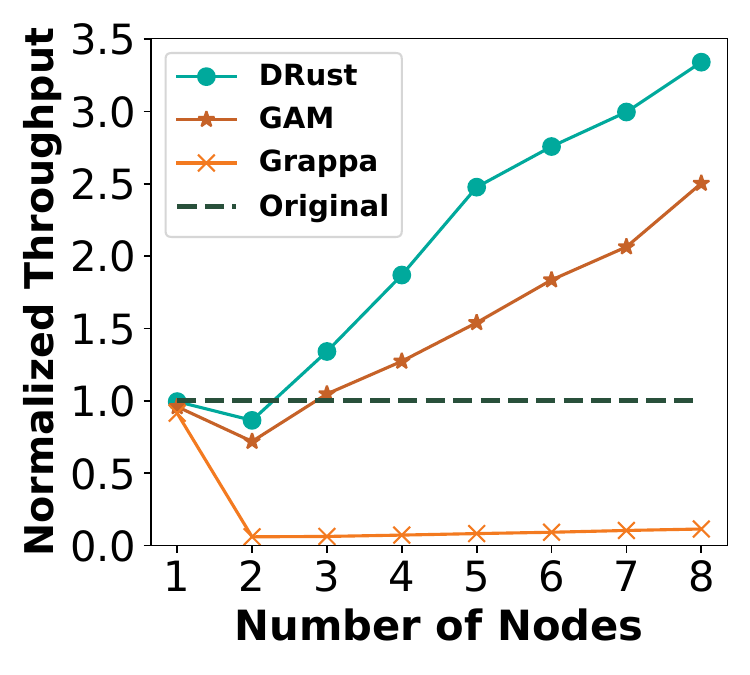}  
         \\
         (a) DataFrame (318 s) &  (b) SocialNet (120 ops/s) &  (c) GEMM (1039 s) & (d) KV Store (2.7 Mops/s)\\
    \end{tabular}
    % \begin{tabular}{cccc}
    %      \includegraphics[width=0.3\linewidth]{figs/dataframe-scale.png}&  \includegraphics[width=0.3\linewidth]{figs/socialnet-scale.png}&  \includegraphics[width=0.3\linewidth]{figs/blas-scale.png}&  \\
    %      (A) Dataframe&  (B) SocialNet&  (C) BLAS&  (D) KV-store\\
    % \end{tabular}
    \vspace{-1em}
    \caption{Application throughput when running with \tool, GAM, and Grappa, normalized to the throughput of their original implementation running on a single node. The number in the parenthesis is the original application's throughput on a single node.\label{fig:overall-performance}}
  \vspace{-1em}
\end{figure*}

\MyPara{DataFrame} is an in-memory data analytics framework similar to Spark~\cite{spark} and Pandas~\cite{pandas}. We built our library atop Polars~\cite{polars}, a native DataFrame engine in Rust offering OLAP query APIs such as \codeIn{filter}, \codeIn{groupby}, and \codeIn{join}. DataFrame organizes the dataset as columnar format tables in shared memory, and user queries will manipulate table columns by reading/writing rows and transforming them into new tables. DataFrame exploits data-level parallelism by internally partitioning columns by row into an array of small chunks where each chunk can be processed independently. We additionally applied \codeIn{TBox} to annotate chunks from the same table column for co-location and used \codeIn{spawn\_to} to offload columnar operations to the data side to improve data locality and performance. 
Note that such annotations were not necessary for the application to run; they were added for additional performance optimizations.
% Our annotations involve 57 lines of code changes to the library. 

\MyPara{SocialNet} is a twitter-like web service from the DeathStarBench suite~\cite{gan2019deathstarbench}. It is composed of 12 microservices with complicated call dependencies. Each microservice in SocialNet can scale independently with replicas, thereby offering higher throughput with more servers. SocialNet decouples the process of user texts, media resources, and storage into different microservices, and it employs RPCs to pass values (texts, media files, \etc) between them. 
\tool enables SocialNet to pass only references in RPCs, eliminating the serialization/deserialization overhead and redundant data copies.
Because SocialNet was implemented in C++ and deployed with Docker Swarm~\cite{dockerswarm}, we ported it into Rust for our evaluation. We followed its original microservice structure but changed the RPC call sites to pass references instead of values, and we followed the original orchestration configuration to spread and scale each microservice in the cluster. We did not use any affinity annotations for SocialNet.
%\yq{Did we use affinity annotations?}\mhr{No affinity annotation for SocialNet}

\MyPara{GEMM} (general purpose matrix multiplication) is a highly-optimized matrix multiplication routine from the BLAS library~\cite{blackford2002blas}. We ported the library using the same divide-and-conquer algorithm by recursively partitioning each matrix into small chunks for parallel processing and reducing the final results. Input and output matrices are stored in the shared memory, where each subroutine will read two input matrix chunks and write the partial results back to the output matrix.
Our port strictly followed the original implementation without using additional affinity annotation.

\MyPara{KV Store} is an in-memory key-value cache engine similar to Memcached~\cite{memcached}. It uses a hash table to store KV pairs in shared memory and mutexes to synchronize concurrent requests. 
We used YCSB benchmark~\cite{cooper2010benchmarking} to generate zipf load with 90\% GET and 10\% SET using default skewness parameter $0.99$. 

\subsection{Scaling Performance}
\label{sec:eval:overall-perf}

In this experiment, we investigated whether \tool can speed up applications by distributing them in a cluster and how well they can scale with the number of servers used. For each application, we first ran it \emph{as is} on a single server without using DSM and measured its throughput. Then, we ran the same application on DSM (subject to modifications when running Grappa) with the same configuration but on varying numbers of servers and measured the throughput normalized to its single-node throughput (\ie, strong scaling). As GAM and Grappa cannot adaptively balance the workload across servers, we evenly distributed the application's working set and threads among all participating nodes.
Ideally, an application should scale linearly and enjoy proportionally higher throughput with more nodes. However, this is usually unachievable because of the limited parallelism of real-world applications and the coherence overhead of DSM systems, and a good result for \tool will show that applications' throughput can get close to their ideal throughput.

Figure~\ref{fig:overall-performance} shows the results for each application respectively. \tool outperforms both GAM and Grappa in all cases. On a single node, it is 1.04--2.10$\times$ faster than two baseline DSMs, while only adding a maximum overhead of 2.42\% compared to the original program. When running with multiple nodes, \tool scales up applications significantly better than GAM and Grappa. On eight nodes, \tool achieves a throughput that is 1.33--2.64$\times$ higher than that of GAM, 2.53--29.16$\times$ higher than that of Grappa.

Compared to each program's single-machine performance, using DSM over \tool enables each program to easily leverage the available distributed resources and achieve a throughput that is 3.34--11.73$\times$ higher than their single-machine counterparts. 
%Note that all applications except SocialNet are originally single-machine applications, so we compared \tool against their single-node throughput; for SocialNet, where the original application can run distributedly, we directly compare \tool against its eight-node throughput. \shan{This sentence is a further explanation of how we compare \tool with the original application when running on eight nodes.} \hx{what does this mean exactly?}
Next, we discuss each application to explain the scalability difference between \tool and the baseline DSMs. 

\MyPara{DataFrame.}
As shown in Figure~\ref{fig:overall-performance}a, compared with its original version, DataFrame running on eight nodes with \tool achieves 5.57$\times$ higher throughput, whereas with GAM and Grappa, the throughput improvements are 2.18$\times$ and 1.69$\times$, respectively. In other words, DataFrame with \tool is 2.56$\times$ and 3.29$\times$ faster than GAM and Grappa on eight nodes, respectively.

A detailed examination reveals that the performance difference comes from the \emph{shared index table} in each DataFrame operation and the \emph{shared chunks} between dependent DataFrame operations.
In each operation, DataFrame constructs an index hash table to track the mapping from each destination chunk in the output column to all its source chunks in the input column. This index table is shared by all index-builder threads and worker threads. During processing, index-builder threads will concurrently insert into the index table using the destination chunk ID as the key and an array of source chunk IDs as the value, and worker threads will look up the shared index table and retrieve source chunks for processing. As a result, the massive writes and reads to the shared table can incur heavy coherence overhead.
Further, DataFrame passes chunks as references between dependent operations and relies on the DSM system for actual data movement. However, it only performs lightweight computation over the fetched data (\ie, low compute intensity as shown in Table~\ref{tab:eval-apps}), making the coherence overhead stand out.

\tool outperforms GAM and scales much better because of its light coherence protocol, which incurs negligible object move overhead for writes and no coherence overhead for reads. The use of affinity annotations also helps DataFrame colocate worker threads with their frequently accessed data, bringing 20\% additional boost (details in \S\ref{sec:eval:drilldown}).
GAM, in contrast, has to invalidate each cache block upon each write and read, thereby bottlenecked by the extensive coherence traffic.
Grappa performs the worst in all three DSM systems due to its \emph{always-delegation} programming model, which implements every global memory read/write via a delegated function call. The cost for delegation overwhelms the actual memory access latency in this case, ruining the performance of the shared hash table. Grappa's delegation overhead actually causes a 1.23$\times$ slowdown when scaling DataFrame from a single node to two nodes.

\MyPara{SocialNet.} Since SocialNet is microservice-based and can be deployed distributively, we added another baseline by running the original (non-DSM) code but deploying it on varying numbers of nodes. Figure~\ref{fig:overall-performance}b demonstrates the performance of all systems. SocialNet runs consistently faster with all three DSM systems compared to the original version. \tool, GAM, and Grappa achieve a 2.18$\times$, 2.02$\times$, and 1.57$\times$ speedup on a single node and a 3.51$\times$, 1.33$\times$, and 1.39$\times$ speedup on eight nodes, respectively.
In the conventional setup, SocialNet requires data\textemdash such as text and media files\textemdash to be serialized into byte streams for network transmission, and then deserialized back into usable formats at the receiving end. This serialization and deserialization process is computationally intensive, particularly for large or complex data objects. In contrast, DSM systems enable SocialNet to pass references instead of the entire data values required by remote procedure calls. This approach eliminates the need for serialization and deserialization, reduces redundant data copies, and significantly enhances performance.
% The speedup over the original version is because DSM enables SocialNet to pass references instead of actual values required by RPCs, eliminating serialization/deserialization and redundant data copies.
\tool scales much better than GAM and Grappa thanks to its lightweight coherence protocol, achieving up to 2.77$\times$ and 3.16$\times$ higher throughput than GAM and Grappa, respectively.

\MyPara{GEMM.} GEMM differs from the previous two applications in its high compute intensity and relatively infrequent shared memory accesses. 
In this application, matrices are transformed and divided into smaller sub-matrices for parallel processing. 
Each computing thread, responsible for multiplying sub-matrices, is assigned to a server. These threads cache their respective sub-matrices in the server's local memory and access them repeatedly to compute product results. This process is highly compute-intensive. As depicted in Figure~\ref{fig:overall-performance}c, \tool and GAM scale well for GEMM and achieve 5.93$\times$, 3.82$\times$ speedup with eight nodes. In contrast, Grappa only achieves a 2.02$\times$ speedup with eight nodes due to its inability to cache sub-matrices locally, necessitating frequent remote accesses.
\tool's superior performance over GAM, with a 1.55$\times$ higher speedup on eight nodes, is primarily due to its more efficient handling of initial cross-server data accesses required when a sub-matrix is first accessed remotely. Unlike GAM, which incurs significant runtime overhead due to the maintenance of state and location of data copies, \tool directly copies data to local memory, without any complex cross-server synchronization operations, thus enhancing overall efficiency.
% \MyPara{GEMM.} GEMM differs from the previous two applications in its high compute intensity and relatively infrequent shared memory accesses. Figure~\ref{fig:overall-performance}c depicts its performance. All three DSM systems scale well for GEMM and achieve 5.93$\times$, 3.82$\times$, and 2.02$\times$ speedup with eight nodes. \tool still outperforms GAM and Grappa by 1.55$\times$ and 2.93$\times$ on eight nodes, respectively, and the performance difference is mainly because of their heavy runtime overhead (for cache coherence or delegation).

\MyPara{KV Store.} KV Store is the most DSM-unfriendly application in our evaluation because it exposes poor memory locality and low compute intensity, which amplifies the overhead of cross-server memory accesses. In addition, it uses mutexes to synchronize between workers and the structure of the program does not lend itself to ownership-based read/write ordering.

Figure~\ref{fig:overall-performance}d shows the results. KV Store experiences a slowdown on all three DSM systems when scaling from a single node to two nodes (13\% for \tool, 25\% for GAM, and 93\% for Grappa). 
However, the impact is mitigated when more servers are enlisted\textemdash \tool and GAM achieve 3.34$\times$ and 2.50$\times$ higher throughput on eight nodes compared to the original KV Store implementation, respectively. 
Due to the limited ownership semantics exposed by mutexes, \tool does not scale as well with KV Store as with other applications. \tool is 1.33$\times$ faster than GAM on eight nodes, benefiting from its adaptive load balancing and a more efficient implementation of mutexes utilizing one-sided RDMA atomic verbs, whereas GAM depends on less efficient two-sided RDMA messages for synchronization. 
Grappa, in contrast, incurs the highest distribution overhead and poorest scalability, primarily because each PUT/GET operation requires remote delegation, and nodes handling popular objects become bottlenecked due to skewed load.

\subsection{Drill-Down Experiments \label{sec:eval:drilldown}}

\begin{figure}
    \begin{minipage}[b]{0.45\linewidth}
        \centering
        \includegraphics[width=1.05\linewidth]{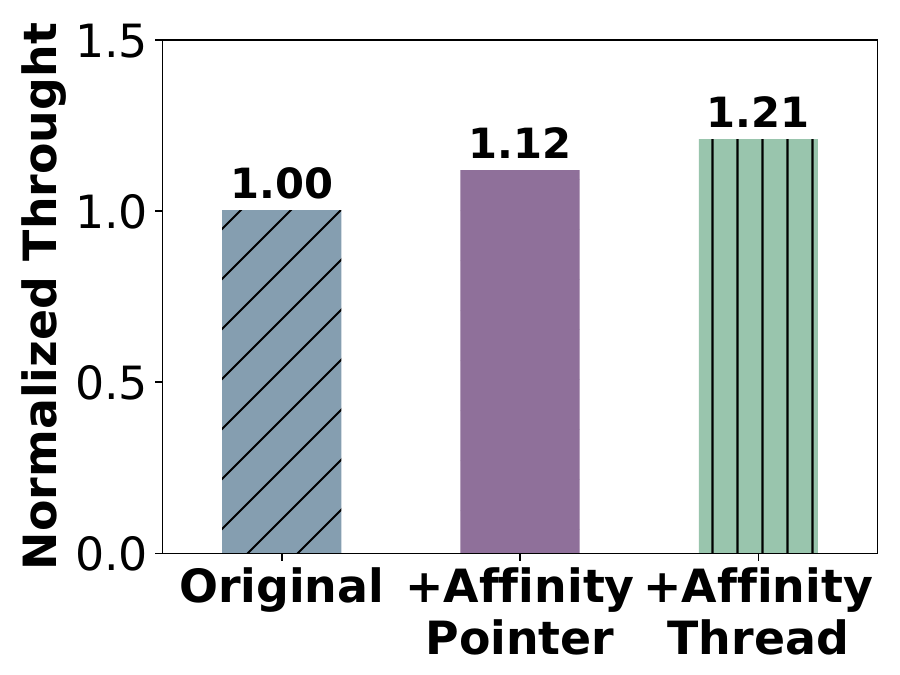}
        % \vspace{-2em}
        \caption{Effectiveness of \tool's affinity annotations.}
        \label{fig:construct-performance}
    \end{minipage}
    \hfill
    \begin{minipage}[b]{0.49\linewidth}
        \begin{adjustbox}{max width=\linewidth}
        \small
        \centering
        \begin{tabular}{c|ccc}
        \toprule
        \thead{Latency\\ (cycles)}  & \thead{Average} & \thead{Median} & \thead{P90} \\\midrule
        \codeIn{\tool} & 395 & 356 & 536 \\
         \codeIn{Rust} & 364 & 332 & 496 \\
        \bottomrule
        \end{tabular}
        \end{adjustbox}
        \captionof{table}{\tool's \codeIn{Box} pointer only adds a small dereferencing cost compared to Rust's ordinary \codeIn{Box}.}
        \label{tab:box-deref-cost}
    \end{minipage}
    \vspace{-1.5em}
\end{figure}

% \begin{figure}
%     \centering
%     \includegraphics[width=0.7\linewidth]{figs/constructs.pdf}
%     \vspace{-1em}
%     \caption{Effectiveness of \tool's affinity annotations.}
%     \label{fig:construct-performance}
%     \vspace{-1em}
% \end{figure}

\MyPara{Affinity Annotations.}
%\label{sec:eval:affinity}
In this experiment, we evaluated the individual contributions of affinity annotations by enabling each of them incrementally for DataFrame on eight nodes. Figure~\ref{fig:construct-performance} reports the results. Using \codeIn{TBox} helps DataFrame group chunks from the same column and eliminates the runtime dereference check overhead for single-column operations (\eg, \codeIn{filter}), bringing a 12\% throughput improvement. Adding \codeIn{spawn\_to} further improves the throughput by 9\% by informing \tool runtime to colocate the worker thread to its input columns, which reduces cross-server memory accesses.

% \tool introduces two additional language constructs: \codeIn{TBox} and \codeIn{spawn\_to} for programmers to use when writing applications to further enhance performance when running on \tool. To demonstrate how much performance improvement we can get through each type of language construct, we evaluated the performance of Dataframe on 8 machines under three different settings: (1) running original Dataframe on \tool without any new language constructs; (2) adding \codeIn{TBox} pointers to co-locate a column’s chunks on a single server and also reduces the amount of runtime location check overhead in dereferencing smart pointers; (3) adding \codeIn{TBox} and also leverage \tool’s \codeIn{spawn\_to} method to schedule single-column operations directly on the server hosting the respective column. The performance comparison is shown in figure~\ref{fig:construct-performance}. It shows that using \codeIn{TBox} brings in 11\% performance improvement and adding \codeIn{spawn\_to} further brings in 9\% more performance benefit.

% \begin{table}[t]
% % \begin{adjustbox}{max width=0.9\linewidth}
%   \small
%     \centering
%     \begin{tabular}{c|ccc}
%     \toprule
%     Latency (cycles)  & Average & Median & P90 \\\midrule
%     \codeIn{\tool} / \codeIn{Rust} & 395 / 364 & 356 / 332 & 536 / 496 \\
%     \bottomrule
%     \end{tabular}
%     \caption{\tool' Box pointer only adds a small dereferencing cost compared to Rust's ordinary Box pointer.}
%     \label{tab:box-deref-cost}
%     \vspace{-2em}
% \end{table}

\MyPara{Runtime Dereference Checks.}
%\label{sec:eval:rt-check}
We measured the latency of dereferencing \tool's \codeIn{Box} pointer and compared it with an ordinary Rust \codeIn{Box} pointer. Both of them point to an 8-byte object in local memory and not in CPU’s cache, which represents the common path for pointer dereferencing. Table~\ref{tab:box-deref-cost} reports the results. \tool only adds a small overhead of $\sim$30 cycles.
% , which is around 8.5\% on average.
Note that this microbenchmark is extremely memory-intensive, whereas real-world applications usually employ larger object sizes and are more compute-intensive, further mitigating the runtime check overhead. For our evaluated applications, we observed a 1.02\% overhead for DataFrame and a 1.14\% overhead for BLAS, when they run with \tool on a single node, respectively.
% When running real-world applications, the object size is usually larger and the application has more computation tasks than just continuously dereferencing pointers, which means the overhead in real-world cases would be much smaller. We evaluate the Dataframe and BLAS on a single machine with \tool's Box pointer and ordinary Rust's Box pointer. \tool shows only 1.02\% overhead for Dataframe and 1.14\% overhead for BLAS.

% \begin{figure}
%     \centering
%     \includegraphics[width=\linewidth]{figs/migration.pdf}
%     \caption{Benefits and overhead of \tool's thread migration technique.}
%     \label{fig:migration}
% \end{figure}

\MyPara{Thread Migration Latency.}
%\label{sec:eval:thd-migration}
To quantify how quickly \tool can resolve the workload imbalance, we measured the latency for the \tool runtime to migrate a thread by running GEMM on eight nodes and repeated the experiment for ten times. On average, \tool migrated 15 threads with an average of 218\us latency for each migration.

\begin{figure}
    \centering
    \includegraphics[width=\linewidth]{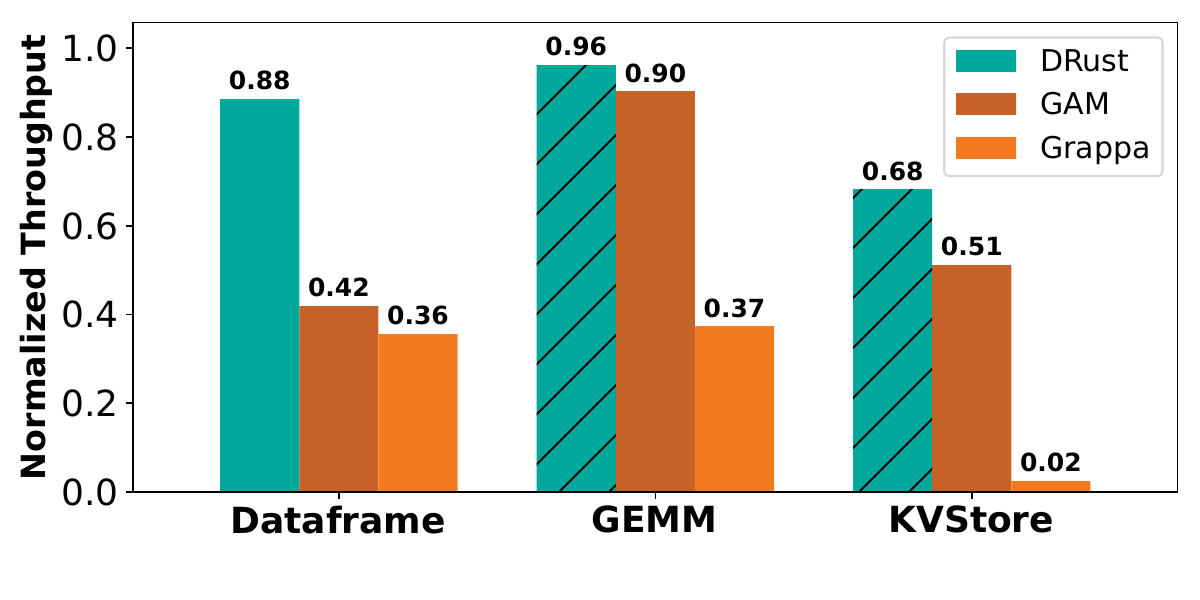}
    \vspace{-3em}
    \caption{Comparison of cache coherence costs between \tool, GAM, and Grappa on eight nodes.}
    \label{fig:coherence-cost}
    \vspace{-1.5em}
\end{figure}

\MyPara{Cost of Cache Coherence.}
%\label{sec:eval:sync-overhead}
In this experiment, we ran each application again on a single node and eight nodes but fixed the total amount of CPU and memory resources. For the eight-node setting, we distributed the resources evenly to each node and measured application throughput. We expect to see a slowdown due to the cost of running the coherence protocol and cross-server memory accesses, but a good result for \tool should show that application performance remains close to its original single-node version. Figure~\ref{fig:coherence-cost} reports the results. SocialNet uses pass-by-value RPCs in its original version and is significantly slower than our DSM-based version, so it is omitted in the evaluation. \tool adds only moderate cache coherence cost with an overhead of 32\% in the worst case (KV Store) and 4\% in the best case (GEMM). GAM and Grappa, in contrast, incur much higher overheads ranging from 10\% to 98\% for different applications. 
\section{Related Work\label{sec:rw}}

\MyPara{Software DSM Systems.}
Distributed cache coherence protocols and their implementations for DSM have been extensively studied since 1980s~\cite{li1988ivy, li1989memory,carter1991implementation,carter1995techniques,stets1997cashmere, minnich1993mether, fleisch1987distributed, campbell1987choices,nieplocha1994global,nieplocha1996global,nieplocha2002combining,gustavson1992scalable}. Among them, Munin~\cite{bennett1990munin} and TreadMarks~\cite{amza1996treadmarks} proposed relaxed consistency models and simpler protocols trying to alleviate the coherence overhead.
Recent DSM systems leveraged today's advanced hardware such as RDMA~\cite{nelson2015latency, kaxiras2015turning, shan2017distributed, cai2018efficient, taranov2021corm, zhang2023partial} to improve efficiency. 
%However, they still run traditional coherence protocols which are inherently expensive.

\MyPara{Disaggregated and Remote Memory.}
Memory disaggregation and remote memory techniques are another promising approach to scaling applications out of a single machine. Their key idea is to connect a host server with large memory pools~\cite{genzconsortium,the-machine-ross15,the-machine-hp} via fast datacenter network, which can be accessed by applications via OS kernel~\cite{legoos-osdi18,fastswap-eurosys20,wang2023canvas,qiao2023hermit} or software runtimes~\cite{ruan2020aifm,semeru@osdi2020,wang2022memliner,mako-pldi22, mira@sosp23,chenxi@jcst23}. However, they do not provide cache coherence.
%and restrict application threads to running on a single server, thereby hindering application scalability.

\MyPara{Distributed Programming Abstractions.}
%To lower the barrier of distributed programming, 
Researchers have studied and proposed new programming languages and abstractions. 
Munin\cite{bennett1990munin} built a type system that defines types for local and global pointers and tracks whether the pointer is shared via type checking. X10~\cite{chandra2008type, haque2015type} and UPC~\cite{el2005upc} introduce function offloading interfaces for distributed computing and additional type annotations to reduce the runtime overhead.
Ray~\cite{rayownership2021} and Nu~\cite{ruan2023nu} are two recent systems proposing new abstractions for distributed programming. Unlike \tool, they require effort to port applications to avoid fine-grained memory sharing. 

\MyPara{Hardware-Accelerated DSM.}
Specialized datacenter network technologies and emerging hardware designs stand for another trend to accelerate DSM. 
Clio \cite{clio@asplos22}, StRoM \cite{strom@eurosys-20}, and RMC \cite{rmc@hotnets20} reduce remote memory access latency by offloading tasks into customized hardware.
Concordia~\cite{wang2021concordia}, Kona~\cite{kona@asplos21}, and CXL~\cite{cxl, cxl1, cxl2, cxl11, zhang2023partial, ms-cxl@arxiv, li2023pond} enable block-level or cache-line-level memory coherence with their hardware-implemented protocols. 
\tool can benefit from advances in hardware support and achieve better scalability.

\section{Conclusion\label{sec:conclusion}}
This paper presents \tool, a practical DSM system based on the ownership model. It automatically turns a single-machine Rust program into its distributed version with a lightweight coherence protocol guided by language semantics. 
\tool significantly outperforms existing state-of-the-art DSM systems, demonstrating that a language-guided DSM can achieve strong memory consistency, transparency, and efficiency simultaneously.

\section*{Acknowledgement}
We thank the anonymous reviewers for their valuable and thorough comments. We are grateful to our shepherd Daniel S. Berger for his feedback. This work is supported by CNS-1763172, CNS-2007737, CNS-2006437, CNS-2106838, CNS-2147909, CNS-2128653, CNS-2301343, CNS-2330831, CNS-2403254, CNS-1764077, CNS-1956322, CNS-2106404. This work is also supported by Alibaba Group through Alibaba Research Intern Program, and funding from Amazon and Samsung.

\begin{comment}
In this paper, we presented \tool, a practical DSM system based on the ownership model. It automatically turns a single-machine Rust program into its distributed version with a lightweight coherence protocol guided by language semantics. \tool significantly outperforms existing state-of-art DSM systems (by X$\times$ in our evaluation), demonstrating that software DSM can achieve strong memory consistency, transparency, and efficiency simultaneously.
\end{comment}
\appendix
% \appendix
\section{Artifact Appendix}
\subsection{Artifact Summary}

\tool is an efficient, consistent, and user-friendly DSM system featuring a lightweight coherence protocol guided by language semantics. \tool allows for seamless scaling of single-machine applications to multi-server environments without sacrificing performance. Demonstrating significant improvements over existing DSM systems, \tool combines strong memory consistency, transparency, and efficiency effectively.

\subsection{Artifact Check-list}
\begin{itemize}
  \item {\bf Hardware:} Intel servers equipped with InfiniBand
  \item {\bf Software Environment:} Rust 1.69.0, GCC 5.5, Linux Kernel 5.4, Ubuntu 18.04, MLNX-OFED 4.9
  \item {\bf Public Link to Repository:} \url{https://github.com/uclasystem/DRust}
  \item {\bf Code License:} GNU General Public License (GPL)
\end{itemize}

\subsection{Description}
\subsubsection{\tool's Codebase} 
\tool comprises four main components: 
\squishlist
\item An RDMA communication library written in C
\item The \tool library
\item Applications integrated with \tool
\item Necessary shell scripts and configuration files
\squishend

\subsubsection{Deploying \tool}  
The initial step in deploying \tool involves cloning the source code on all involved servers:
\begin{mdframed}[style=MyFrame,nobreak=true]
\codeIn{git clone git@github.com:uclasystem/DRust.git}
\end{mdframed}

Adjust several configurations according to your server setup and operational requirements:
\begin{enumerate}
    \item Set the Number of Servers:
    \begin{itemize}
        \item Define \codeIn{TOTAL\_NUM\_SERVERS} in \codeIn{comm-lib/rdma-common.h} based on the total number of available servers.
        \item Similarly, adjust \codeIn{NUM\_SERVERS} in \codeIn{drust/src/conf.rs}.
    \end{itemize}
    \item Configure the Distributed Heap Size by setting \codeIn{UNIT\_HEAP\_SIZE\_GB} in \codeIn{drust/src/conf.rs} to the required heap size per server, e.g., 16 for 16GB.
    \item Update the InfiniBand IP addresses and ports in \codeIn{comm-lib/rdma-server-lib.c}:
    \begin{mdframed}[style=MyFrame,nobreak=true]
    \codeIn{const char *ip\_str[2] = \{"10.0.0.1", "10.0.0.2"\};}\\
    \codeIn{const char *port\_str[2] = \{"9400", "9401"\};}
    \end{mdframed}
    \item In \codeIn{drust.json}, update each server's IP address and specify three available ports.
\end{enumerate}

% Several configurations need to be adjusted based on your server setup and requirements. Follow these steps to configure \tool:
% \begin{enumerate}
%     \item Set the Number of Servers:
%     \begin{itemize}
%         \item In \codeIn{comm-lib/rdma-common.h}, set \codeIn{TOTAL\_NUM\_SERVERS} to the total number of servers you have.
%         \item In \codeIn{drust/src/conf.rs}, set \codeIn{NUM\_SERVERS} to the same number.
%     \end{itemize}
%     \item Configure Distributed Heap Size. In \codeIn{drust/src/conf.rs}, set \codeIn{UNIT\_HEAP\_SIZE\_GB} to the desired heap size for each server (e.g., 16 for 16GB).
%     \item Set InfiniBand IP Addresses and Ports. In \codeIn{comm-lib/rdma-server-lib.c}, update \codeIn{ip\_str} and \codeIn{port\_str} arrays with your servers' IP addresses and ports. Example:
%     \begin{mdframed}[style=MyFrame,nobreak=true]
%     \codeIn{const char *ip\_str[2] = \{"10.0.0.1", "10.0.0.2"\};}\\
%     \codeIn{const char *port\_str[2] = \{"9400", "9401"\};}
%     \end{mdframed}
%     \item Configure Server IP Addresses in \codeIn{drust.json}. Modify \codeIn{drust/drust.json} with each server's IP address and three available ports.
% \end{enumerate}

Following configuration, build \tool as follows:
\begin{mdframed}[style=MyFrame,nobreak=true]
\codeIn{\# Compile the communication static library}\\
\codeIn{cd comm-lib}\\
\codeIn{make -j lib}\\
\codeIn{\# Copy the static library to the \tool directory}\\
\codeIn{cp libmyrdma.a ../drust/}\\
\codeIn{\# Compile the Rust components}\\
\codeIn{cd ../drust}\\
\codeIn{cargo build \textendash\textendash release}
\end{mdframed}

Deploy the compiled binary across all servers post-build, ensuring its correct distribution:
\begin{mdframed}[style=MyFrame,nobreak=true]
\codeIn{scp target/release/drust user@ip:DRust/drust.out}
\end{mdframed}

\subsubsection{Running Applications}

\tool is bundled with four example applications: Dataframe, GEMM, KVStore, and SocialNet. Follow these steps to execute them:
\begin{enumerate}
    \item Launch the \tool executable on all servers, excluding the main server:
    \begin{mdframed}[style=MyFrame,nobreak=true]
\codeIn{\# Start the \tool process with the specified server index and application name.}\\
\codeIn{\# For example, ./../drust.out -s 7 -a gemm}\\
\codeIn{cd drust}\\
\codeIn{./../drust.out -s \textit{server\_id} -a \textit{app\_name}}
\end{mdframed}
    \item On the main server:
        \begin{mdframed}[style=MyFrame,nobreak=true]
\codeIn{\# Start the main \tool process with the specified application.}\\
\codeIn{cd drust}\\
\codeIn{./../drust.out -s 0 -a \textit{app\_name}}
\end{mdframed}
\end{enumerate}

More details of \tool's installation and deployment can be found in \tool's code repository.

\newpage
% \appendix
\section{Full Coherence Protocol \label{sec:full-design}}

This appendix provides a full version of our coherence protocol, on which our proof of memory coherence is conducted. It includes more detailed descriptions of the algorithms, specifically Algorithms~\ref{alg:imm-appendix}, \ref{alg:mut-appendix}, \ref{alg:ownerimm-appendix}, and~\ref{alg:ownermut-appendix}, which illustrate the dereferencing and dropping procedures for each type of reference. 
Additionally, Algorithm \ref{alg:util-appendix} describes various utility functions that support these operations.
% Algorithm \ref{alg:drop} explains the procedures for safely dropping references and owner \codeIn{Box} pointers. 

\subsection{Pointer Layout}

% As discussed in the paper, each pointer in \tool must remember not only the object's global address, but also the address of the cached copy in a server's local cache (to avoid redundant remote fetches). As such, we modify Rust's pointer structure, as illustrated in Figure~\ref{fig:ptr-layout}. First, \tool internally extends each Rust \codeIn{Box} pointer and reference with an additional 64-bit field, which is used differently for read and write access. And in the global address field, we reserve 16 tag bits for recording the version of the value pointed by the reference. We also reserve 1 bit from the second field to record if the tag has been updated or not. As you will see from our detailed algorithms, the tag field will be updated every time a mutable reference is firstly dereferenced or the owner is firstly doing write access after all immutable references return. In this way, newly created immutable references always get a new object address (or a new tag) if the object has been modified, so that it will read deprecated cache copies in its local cache.
\tool extends each Rust \codeIn{Box} pointer and reference with an additional 64-bit field, as depicted in Figure~\ref{fig:full-ptr-layout}. Additionally, in the global address field, we reserve 16 bits as ``color'' bits, which are used to track the version of the referenced value. Every time a mutable reference is dereferenced for the first time, or the owner performs a write operation subsequent to the return of all immutable references, the color bits are updated. Consequently, this mechanism guarantees that any newly created immutable references will obtain a new global address\footnote{In the following text, a ``colored global address'' means the composition of a global address and its color.}. This update is critical for ensuring that outdated cache copies in the local cache are not utilized, thereby guaranteeing the coherence of \tool. Moreover, we repurpose one bit in the extension field to indicate whether the color has been updated. This design choice reduces the frequency of color updates during continuous local writes by the owner, thereby minimizing the frequency of color overflow.

\subsection{Immutable Reference}
\begin{figure}
    \centering
    \includegraphics[width=0.95\linewidth]{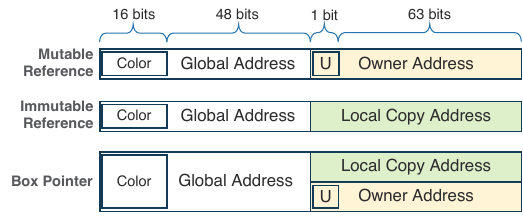}
    \caption{\tool's pointer layout. \label{fig:full-ptr-layout}}
\end{figure}

The procedure for dereferencing an immutable reference is outlined in Algorithm~\ref{alg:imm-appendix}. This function accepts an immutable reference, consisting of a colored global address and a local copy address (see Figure~\ref{fig:full-ptr-layout}), and outputs a local memory address for direct reading. The initial step involves determining the location of the object using the \codeIn{IsLocal} function (Line~\ref{l:islocal-immut-a}). If this address is local, \tool clears the color embedded in the global address and returns it for direct access (Line~\ref{l:cleartag-immut-a}).

For a remote object, \tool checks whether the local copy field is \codeIn{null} (Line~\ref{l:check-copy-null-a}). If this field is \codeIn{null}, indicating no recorded local cache copy, the algorithm proceeds to examine the cache hashmap $H$ on the server executing the access. This hashmap, keyed by the colored global address, contains entries pairing local copy addresses with a count of local immutable references pointing to this copy. The algorithm checks for the presence of the colored global address $r.g$ in $H$ (Line \ref{l:checkcache-immut-a}). If found, it increments the reference count in that entry and updates the local copy field in the immutable reference with the local address from the entry (Lines~\ref{l:cachehit-immut-st-a}-\ref{l:cachehit-immut-ed-a}). If $r.g$ is not in $H$, the object is copied from the remote server, stored locally, and the local address is assigned to $r.l$ (Line~\ref{l:cachemiss-immut-st-a}). This newly fetched object, along with its associated reference count, is then added to $H$ (Line~\ref{l:cachemiss-immut-ed-a}). Both the update and lookup operations on $H$ are performed atomically to avoid concurrency issues.

The procedure for dropping an immutable reference is outlined in the \codeIn{DropRef} function (Lines~\ref{l:droprefst-a}-\ref{l:droprefed-a}) in Algorithm~\ref{alg:imm-appendix}. It checks the local copy address field of the reference. If it is not \codeIn{null}, \tool decrements the reference count for this local copy in $H$. Once the reference count hits zero, the local copy becomes eligible for eviction by \tool's runtime system.

\begin{algorithm}[t]
\small
\KwIn{A shared immutable reference $r$ containing a global address $r.g$ and a local copy address $r.l$, and a local cache hashmap $\mathit{H}$. }
\KwOut{A local memory address for reading.}
\DontPrintSemicolon
\SetKwFunction{Derefmut}{\textsc{DerefMut}}
\SetKwFunction{Dropmutref}{\textsc{DropMutRef}}
\SetKwFunction{Getcolor}{\textsc{GetColor}}
\SetKwFunction{Clearcolor}{\textsc{ClearColor}}
\SetKwFunction{Appendcolor}{\textsc{AppendColor}}
\SetKwFunction{Islocal}{\textsc{IsLocal}}
\SetKwFunction{Move}{\textsc{Move}}
\SetKwFunction{Copy}{\textsc{Copy}}
\SetKwFunction{Getentry}{\textsc{GetEntry}}
\SetKwFunction{Updateentry}{\textsc{UpdateEntry}}
\SetKwFunction{Insertentry}{\textsc{InsertEntry}}
\SetKwFunction{Write}{\textsc{Write}}
\SetKwFunction{Deref}{\textsc{Deref}}
\SetKwFunction{Dropref}{\textsc{DropRef}}

    \SetKwProg{Fn}{Function}{:}{}
    \Fn{\Deref{$r$, $H$}}{
        \If{\Islocal{$r.g$}}{ \label{l:islocal-immut-a}
            \Return \Clearcolor{$r.g$} \label{l:cleartag-immut-a}\;
        }
        \Else{
            \If{$l = \mathit{Null}$}{\label{l:check-copy-null-a}
                \textsc{\textbf{Atomic}} \{ \\
                    \If{$r.g \in H$}{\label{l:checkcache-immut-a}
                        $\langle l', cnt \rangle \gets$ \Getentry{$H$, $r.g$}\label{l:cachehit-immut-st-a}\;
                        $r.l \gets l'$\;
                        \Updateentry{$H$, $r.g$, $\langle l'$, $cnt+1 \rangle$}\label{l:cachehit-immut-ed-a}\;
                    }
                    \Else{\label{l:cache-exists-a}
                        $r.l \gets$ \Copy{\Clearcolor{$r.g$}} \label{l:cachemiss-immut-st-a}\;
                        \Insertentry{$H$, $r.g$, $\langle r.l, 1 \rangle$}\label{l:cachemiss-immut-ed-a}\;
                    }
                \}\\

                }
            \Return $r.l$\;
        }
    }

    \Fn{\Dropref{$r$, $H$}}{\label{l:droprefst-a}
        \If{$r.l \neq \text{Null}$}{
            \textsc{\textbf{Atomic}} \{ \\
            $\langle l'$, $cnt \rangle \gets$ \Getentry{$H$, $r.g$}\;
            \Updateentry{$H$, $r.g$, $\langle l'$, $cnt-1 \rangle$}\label{l:drop-ref-counter-a}\;
            \}\\
        }
    }\label{l:droprefed-a}
\caption{Access logic for immutable reference.\label{alg:imm-appendix}}
\end{algorithm}

\begin{algorithm}[t]
\small
\DontPrintSemicolon
\SetKwFunction{Colorupdated}{\textsc{ColorUpdated}}
\SetKwFunction{Clearubit}{\textsc{ClearUBit}}
\SetKwProg{Fn}{Function}{:}{}

\Fn{\Getcolor{$g$}}{
    \Return $g \gg 48$\;
}

\Fn{\Clearcolor{$g$}}{
    \Return $g$ \texttt{\&} $((1 \ll 48) - 1)$\;
}

\Fn{\Appendcolor{$g$, $c$}}{
    \Return \Clearcolor{$g$} \texttt{|} ($c \ll 48$)\;
}

\Fn{\Colorupdated{$o$}}{
    \Return $(o \gg 63)$ \texttt{\&} $1$\;
}

\Fn{\Clearubit{$o$}}{\label{l:clearubit}
    \Return $o$ \texttt{\&} $((1 \ll 63) - 1) $\;
}\label{l:updatetag-ed}

\caption{Utility functions for pointer coloring.\label{alg:util-appendix}}
\end{algorithm}

\subsection{Mutable Reference}

\begin{algorithm}[t]
\small
\KwIn{A mutable reference $m$ containing a global address $m.g$ and the owner address $m.o$. }
\KwOut{A local memory address to be written to.}
\DontPrintSemicolon
% \SetKwFunction{Main}{\textsc{DerefMut}}
%     \SetKwProg{Fn}{Function}{:}{}
%     \Fn{\Main{$\langle g, o \rangle$}}{
%         \If{\textsc{IsLocal}($g$)}{
%             \If{$\neg$\textsc{TagUpdated}($o$)}{\label{l:checktag-mut}
%                 $\langle g, o \rangle \gets \textsc{UpdateTag}(g, o)$\label{l:update-tag-mut}\;
%             }
%         }
%         \Else{
%             $tag \gets \textsc{GetTag}(g)$\;
%             $g \gets \textsc{Move}(\textsc{ClearTag}(g))$\; \label{l:move-mut}
%             $g \gets \textsc{AppendTag}(g, tag)$\;
%         }
%         \Return \textsc{ClearTag}(g)\;
%     }
    
    \SetKwProg{Fn}{Function}{:}{}
    \Fn{\Derefmut{$m$}}{
        \If{\Islocal{$m.g$}}{ \label{l:place-check-a}
            \If{$\neg$\Colorupdated{$m.o$}}{\label{l:checktag-mut-a}
                $m.o \gets m.o$  \texttt{|} $(1 \ll 63)$  \label{l:update-tag-mut-st-a}\;
                $c' \gets$ \Getcolor{$m.g$} + 1 \label{l:tag-increment-a}\;
                $m.g \gets$ \Appendcolor{$m.g$, $c'$} \label{l:update-tag-mut-ed-a}\;
            }
        } \Else{
            % $c \gets$ \Getcolor{$m.g$}\; \label{l:move-st}
            $m.o \gets m.o$  \texttt{|} $(1 \ll 63)$ \;
            $m.g \gets$ \Move{\Clearcolor{$m.g$}}\; \label{l:move-mut-a}
            % $m.g \gets$ \Appendcolor{$m.g$, $c$}\; \label{l:move-ed}
        }
        \Return \Clearcolor{$m.g$}\;
    }

    \Fn{\Dropmutref{$m$}}{ \label{l:dropmutst-a}
    $owner \gets$ \Clearubit{$m.o$}\;
    \Write{$owner$, $m.g$}\; \label{l:writeback-mut-a}
    }\label{l:dropmuted-a}
\caption{Access logic for mutable references.\label{alg:mut-appendix}}
\end{algorithm}

% For objects outside the local heap partition, \tool executes a \codeIn{move} operation (Line~\ref{l:move-mut}), which includes copying the object to a local address, updating the mutable reference’s global address, and instructing the remote server to release the original object.

% Upon expiration of a mutable reference, demonstrated in the \codeIn{DropMutRef} function (Lines~\ref{l:dropmutst}-\ref{l:dropmuted}), \tool records the mutable reference's colored global address back to the owner’s global address field. This ensures the owner’s view remains current, and that any modifications via mutable references on the same machine as the owner are reflected, updating the cache key for future immutable references and preventing stale data access.

Algorithm~\ref{alg:mut-appendix} outlines the logic for dereferencing a mutable reference consisting of a colored global address and the owner's address (see Figure~\ref{fig:full-ptr-layout}). It returns a local memory address for writing. The first step in the algorithm is to determine the location of the referenced object (Line~\ref{l:place-check-a}). If the object is in the local heap partition, \tool updates the color field in the mutable reference. This update is crucial for maintaining data integrity and coherence.
A critical aspect of this mechanism is the first bit of the extension field, the \textit{U} bit, which indicates whether the mutable reference's color has been updated. The algorithm checks the \textit{U} bit (Line~\ref{l:checktag-mut-a}); if it has not been updated, \tool increments the color field and sets the \textit{U} bit to 1 (Lines~\ref{l:update-tag-mut-st-a}-\ref{l:update-tag-mut-ed-a}) . This process prevents redundant color updates during subsequent dereferences, thereby enhancing efficiency.

For remote objects, \tool moves them to the local heap partition (Line~\ref{l:move-mut-a}). The \codeIn{move} operation involves several steps: copying the object to a local address, updating the global address field in the mutable reference to this new address, and requesting the remote server to deallocate the original object.

Upon expiration of a mutable reference, as demonstrated in the \codeIn{DropMutRef} function (Lines~\ref{l:dropmutst-a}-\ref{l:dropmuted-a}), \tool writes the mutable reference's colored global address back to the owner’s global address field. This ensures that the owner maintains an up-to-date view of the data object. Additionally, writing back the updated color guarantees that data modifications through a mutable reference on the same machine as the owner can also be reflected in owner's global address field. This reflection changes the key used for cache look-ups by subsequent immutable references, thereby preventing access to stale data copies cached on other servers.

% This step is necessary because the relocation operation renders the original owner's \codeIn{Box} a dangling pointer, pointing to an invalid memory address. Due to the single-writer invariant, when the mutable reference is active, no other entity, including the owner, can access the data. Therefore, \tool can safely update the owner's global address field to the new data address when the mutable reference is dropped. 

\subsection{Owner}
\label{sec:owner}

% \begin{algorithm}[t]
% \small
% \DontPrintSemicolon
% \SetKwProg{Fn}{Function}{:}{}
% \Fn{\textsc{DropRef}($\langle g, l \rangle, H$)}{\label{l:droprefst}
%     \If{$l \neq \text{Null}$}{
%         $\langle l, c \rangle \gets$ \textsc{GetEntry}$(g)$\;
%         \textsc{UpdateEntry}$(g, \langle l, c-1 \rangle)$\;
%     }
% }\label{l:droprefed}

% \Fn{\textsc{DropMutRef}$(\langle g, o \rangle$)}{ \label{l:dropmutst}
%     $owner \gets$ \textsc{ClearUpdatedBit}$(o)$\;
%     \textsc{Write}$(owner, g)$\; \label{l:writeback-mut}
% }\label{l:dropmuted}

% \Fn{\textsc{Drop}($\langle g, l\rangle$, $\mathit{H}$)}{\label{l:dropownerst}
%     \If{$l \neq \text{Null}$}{
%         \textsc{RemoveEntry}$(H, g)$\;
%         \textsc{Deallocate}$(l)$\;
%     }
%     \textsc{Drop}$(\textsc{ClearTag}(g))$\;
% }\label{l:dropownered}

% \caption{Drop functions.\label{alg:drop}}
% \end{algorithm}

\begin{algorithm}[t]
\small
\KwIn{A \codeIn{Box} pointer $p$ containing a global address $p.g$ and a local copy address $p.l$, and a local cache hashmap $\mathit{H}$. }
\KwOut{A local memory address for reading}
\DontPrintSemicolon
    \SetKwProg{Fn}{Function}{:}{}
    \Fn{\Deref{$p$, $H$}}{
        \If{\Islocal{$p.g$}}{ \label{l:local-immut-owner-st-a}
            \Return \Clearcolor{$p.g$} \label{l:cleartag-immut-onwer-a}\;
        }\label{l:local-immut-owner-ed-a}
        \Else{\label{l:remote-immut-owner-st-a}
            \If{$l = \mathit{Null}$}{
                \textsc{\textbf{Atomic}} \{ \\
                    \If{$p.g \in H$}{
                        $\langle l', cnt \rangle \gets$ \Getentry{$H$, $p.g$}\;
                        $p.l \gets l'$\;
                        \Updateentry{$H$, $p.g$, $\langle l'$, $cnt+1 \rangle$}\;
                    }
                    \Else{\label{l:cache-exists-owner-a}
                        $p.l \gets$ \Copy{\Clearcolor{$p.g$}}\;
                        \Insertentry{$H$, $p.g$, $\langle p.l, 1 \rangle$}\;
                    }
                \}\\

                }
            \Return $p.l$\;
        }\label{l:remote-immut-owner-ed-a}
    }
\caption{Immutable access logic for owner. \label{alg:ownerimm-appendix}}
\end{algorithm}

\begin{algorithm}[t]
\small
\KwIn{A \codeIn{Box} pointer $p$ containing a global address $p.g$ and a local copy address $p.l$, and a local cache hashmap $\mathit{H}$. }
\KwOut{A local memory address to be written to}
\DontPrintSemicolon
\SetKwFunction{Drop}{\textsc{Drop}}
\SetKwFunction{Deallocate}{\textsc{Deallocate}}
\SetKwFunction{Removeentry}{\textsc{RemoveEntry}}
\SetKwFunction{Asyncinvalidate}{\textsc{AsyncInvalidate}}
% \SetKwFunction{Main}{\textsc{DerefMut}}
    \SetKwProg{Fn}{Function}{:}{}
    \Fn{\Derefmut{$p$, $\mathit{H}$}}{
        \If{\Islocal{$p.g$}}{
            \If{$\neg$\Colorupdated{$p.o$}}{\label{l:checktag-mut-owner-a}
                $p.o \gets p.o$  \texttt{|} $(1 \ll 63)$  \label{l:update-tag-mut-owner-st-a}\;
                $c' \gets$ \Getcolor{$p.g$} + 1 \label{l:tag-owner-increment-a}\;
                $p.g \gets$ \Appendcolor{$p.g$, $c'$} \label{l:update-tag-mut-owner-ed-a}\;
            }
        }
        \Else {
            \If{$p.g \notin H$}{\label{l:cache-check-owner-st-a}
                % $c \gets$ \Getcolor{$p.g$}\; \label{l:move-owner-st-a}
                $p.o \gets 1 \ll 63$ \;
                $p.g \gets$ \Move{\Clearcolor{$p.g$}}\; \label{l:move-owner-a}
                % $p.g \gets$ \Appendcolor{$p.g$, $c$}\; \label{l:move-owner-ed-a}
            }
            \Else{\label{l:cachehit-mut-owner-st-a}
                $\langle l$, $cnt \rangle \gets$ \Getentry{$H$, $p.g$}\;
                \Removeentry{$H$,$p.g$}\;
                \Deallocate{\Clearcolor{p.g}}\;
                % $c \gets$ \Getcolor{$p.g$}\;
                % $p.g \gets$ \Appendcolor{$l$, $c$} \label{l:update-g-owner-a}\;
                $p.g \gets l$ \label{l:update-g-owner-a}\;
                $p.o \gets 1 \ll 63$\;
            }\label{l:cachehit-mut-owner-ed-a}
        }
        \Return \Clearcolor{$p.g$}\;
    }

% \Fn{\Drop{$p$}}{\label{l:dropownerst}
%     \Asyncinvalidate{\Clearcolor{p.g}}\;
%     \Drop{\Clearcolor{p.g}}\;
% }\label{l:dropownered}
\caption{Mutable access logic for owner. \label{alg:ownermut-appendix}}
\end{algorithm}

% In \tool, data access via an owner can be mutable or immutable. At runtime, while the access type (mutable or immutable) is deterministic, the presence of concurrent references during immutable access remains unknown. Consequently, akin to immutable references, the \codeIn{Box} pointer adopts the same dual-address layout as immutable reference. We now talk about how to handle immutable (read) access through the owner and mutable (write) access through the owner.
In \tool, memory access through an owner can be mutable or immutable. The access type can be determined at runtime, but the presence of concurrent immutable references during immutable access is not always known. Therefore, similar to immutable references, the \codeIn{Box} pointer for an owner employs a dual-address layout. Here we discuss the handling of both immutable (read) and mutable (write) access through the owner.

\begin{itemize}
    \item \textbf{Immutable Access.}
    As shown in Algorithm~\ref{alg:ownerimm-appendix}, this process begins with a check of the object location. If it is in local heap partition, the global address is directly returned for dereferencing (Lines~\ref{l:local-immut-owner-st-a}-\ref{l:local-immut-owner-ed-a}). If not, \tool emulates the handling for immutable references (Lines~\ref{l:remote-immut-owner-st-a}-\ref{l:remote-immut-owner-ed-a}). The algorithm checks if the local copy address field is \codeIn{null}. If it is, \tool checks the hashmap for existing local copies, potentially copying data to local memory, updating the hashmap, and recording the local address in the local copy field. Otherwise, \tool directly dereferences this local address.
    \item \textbf{Mutable Access.}
    As detailed in Algorithm~\ref{alg:ownermut-appendix}, writing objects through the owner begins by checking the object's location. If it is in local heap partition and the \textit{U} bit is 0 (Line~\ref{l:checktag-mut-owner-a}), the owner's color will be updated and the \textit{U} bit will be set (Lines~\ref{l:update-tag-mut-owner-st-a}--\ref{l:update-tag-mut-owner-ed-a}). If the object is remote, instead of directly moving the object to local memory, this algorithm first checks the hashmap $H$ for a local copy (Line~\ref{l:cache-check-owner-st-a}). If no local copy exists, \tool mimics the mutable reference dereferencing, moving the object to a local address and updating the owner's global address field (Line~\ref{l:move-owner-a}). If a local copy is found, \tool retrieves this local address from $H$ and sets the global address field of the mutable reference to this local address, while also deallocating the original remote copy (Lines~\ref{l:cachehit-mut-owner-st-a}--\ref{l:cachehit-mut-owner-ed-a}).
\end{itemize}

Furthermore, when an owner goes out of scope, \tool mimics the behavior of Rust’s standard dropping function for a \codeIn{Box} pointer. It first drops every object owned by the referenced object, and then deallocates the object pointed to by this \codeIn{Box} pointer. Importantly, the deallocation process in \tool during moving or dropping an object also asynchronously invalidates all cached copies of the object across all servers, ensuring that references to newly allocated objects at the same address do not retrieve stale cached data.

Additionally, it is crucial to reset the $U$ bit in the extension field of the owner or a mutable reference when creating immutable references from them. This ensures that any subsequent data mutation\textemdash occurring after all immutable references have returned\textemdash can update the color again. Without this reset, subsequent immutable references following a data mutation could potentially access stale local copies.

% invokes the \codeIn{Drop} function as shown in Algorithm~\ref{alg:drop} (lines~\ref{l:dropownerst}-\ref{l:dropownered}). This function checks the local copy address field of the \codeIn{Box} pointer; if it's not null (indicating the existence of a local cached copy), \tool deallocates this local copy and removes the corresponding entry from the local cache hashmap to prevent memory leakage. Subsequently, the data pointed to by the global address field of \codeIn{Box} pointer is dropped, after which the owner itself can be safely deallocated.

\section{Memory Coherence Proof\label{sec:proof}}

% \revise{Sequential consistency necessitates a coherent memory system, requiring not only the SWMR invariant but also the \emph{data-value} invariant~\cite{prime-coherence-book2020}.} In simple terms, the data-value invariant requires that the latest write to a value is immediately visible to subsequent readers. As discussed earlier, \tool's protocol moves an object upon a write and updates the owner immediately. Therefore, the latest value is globally visible after each mutable borrow finishes. Subsequent read accesses, either in the Owned state or the Shared state, are hence guaranteed to see the moved object and read its latest value. 

We prove the memory coherence of \tool by demonstrating its adherence to two critical invariants as outlined in \cite{prime-coherence-book2020}.

\begin{itemize}
    \item \textbf{Single-Writer-Multiple-Reader (SWMR) Invariant.} This invariant stipulates that for any specific memory location, there is either a singular writer permitted to write to it or several readers that may read from it at any given time.
    \item \textbf{Data-Value Invariant.} This invariant asserts that if a memory location is read, it must contain the most recent value updated by the writers. 
\end{itemize}

Given that the SWMR invariant is inherently upheld by Rust's ownership-based type system, our focus shifts to proving that our algorithm guarantees the \textit{Data-Value} Invariant. To accomplish this, we will demonstrate the following four invariants of our coherence protocol. Collectively, these invariants ensure the \textit{Data-Value} invariant, thereby substantiating that our system achieves memory coherence in a distributed environment.

\subsection{Atomicity of Global-Address Change}
\vspace{.5em}
\begin{myquote} 
\begin{mdframed}
{\small
\setlength{\abovedisplayskip}{0pt}%
\underline{Atomicity Invariant}: 
The global address, including the color component, stored in the \codeIn{Box} pointer and mutable references can only be modified by a single entity at any given time.
}
\end{mdframed}
\end{myquote}

% There are only 3 scenarios that the global address including the tag part will be changed in our system.
In \tool, the colored global address is subject to modification in three specific scenarios:
\begin{enumerate}
    \item Dereferencing a mutable reference. This action modifies either the global address itself or the color in the global address field of a mutable reference  (Line~\ref{l:update-tag-mut-st-a}--\ref{l:update-tag-mut-ed-a} and~\ref{l:move-mut-a} in Algorithm~\ref{alg:mut-appendix}).
    % Dereferencing a mutable reference which changes its own global address or the tag in the global address field
    \item Mutable dereferencing of the owner. This operation either alters the color or updates the global address to point to a local address (Lines~\ref{l:update-tag-mut-owner-st-a}--\ref{l:update-tag-mut-owner-ed-a},~\ref{l:move-owner-a},~\ref{l:update-g-owner-a} in Algorithm~\ref{alg:ownermut-appendix}).
    % Mutable dereferencing the owner which changes the tag or modifies the global address to let it point to local memory location.
    \item Dropping a mutable reference. This scenario involves updating the owner's global address field with the colored global address stored in the mutable reference (Line~\ref{l:writeback-mut-a} in Algorithm~\ref{alg:mut-appendix}).
    % Dropping a mutable reference which will write its own global address field including the tag to the owner's global address field.
\end{enumerate}

% Scenarios 1 and 2, and scenarios 1 and 3 are changing different contents (one changes global address field of \codeIn{Box} pointer and one changes that in mutable references) so they will never be mutating the same global address at the same time. For scenarios 2 and 3, one possible case is a dropping a mutable reference created from the owner \codeIn{Box} pointer. But in this case, the ownership model's singular write propery guarantees that no access is allowed though owner when a mutable reference is still alive, so these two scenarios will not happen at the same. So the changes to global address (including tag) is guaranteed to be atomic. 
The first two scenarios impact different types of pointers (one changes the global address field of mutable references, and the other changes that of \codeIn{Box} pointers), ensuring they do not concurrently mutate the same memory location. The same case for scenario 1 and 3. Regarding the interaction between the scenario 2 and 3, a potential overlap could occur when dropping a mutable reference created from a owner \codeIn{Box} pointer. However, the ownership model's single-writer property ensures that the access through the owner is prohibited while a mutable reference is alive, preventing these scenarios from happening simultaneously. Therefore, we can conclude that modifications to each colored global address are guaranteed to be atomic.

\subsection{Global-Address-Change-on-Write Invariant}
\vspace{.5em}
\label{sec:acow-inv}
\begin{myquote} 
\begin{mdframed}
%\vspace{-.5em}
{\small
\setlength{\abovedisplayskip}{0pt}%
\underline{Global-Address-Change-on-Write Invariant}: The colored global address of an object always changes following modifications made by either the owner or a mutable reference.
}
\end{mdframed}
\end{myquote}

% Firstly, modifying data that is not in local heap partition guarantees the change of global address because data has to be moved between servers. For modifying data that is already in local heap partition, as can be seen from line in algorithm~\ref{alg:mut} and algorithm~\ref{alg:ownermut}, the tag in the global address will be updated. And when dropping a mutable reference, the updated tag and global address will also be written back to the owner, guarantee the owner has the changed global address after the mutable reference borrowing from it changes the data. 
Modifications to data not present in the local heap partition inherently result in a change of the global address, as the data must be moved between servers. For data already in the local heap partition, as delineated in Lines~\ref{l:update-tag-mut-st-a}--\ref{l:update-tag-mut-ed-a} of Algorithm~\ref{alg:mut-appendix} and Lines~\ref{l:update-tag-mut-owner-st-a}--\ref{l:update-tag-mut-owner-ed-a} in Algorithm~\ref{alg:ownermut-appendix}, the color in the global address field is updated. Moreover, when a mutable reference is dropped, the updated colored global address is written back to the owner (Line~\ref{l:writeback-mut-a} in Algorithm~\ref{alg:mut-appendix}), ensuring that the owner possesses an updated colored global address after the mutable reference alters the data.

% The only complicated part in this property is the \textit{U} bit we have to save some update operations to the tag. Every time an immutable reference is created through the owner or the mutable reference, that entity's \textit{U} bit will be reset to 0 (as discussed at the end of \S\ref{sec:owner}). And next time when we do mutable dereferencing to a reference or \codeIn{Box} pointer, the tag in that entity will be changed. We can do this because actually the access to a data can be divided into epoch in Rust's ownership model. In each epoch, if immutable reference exists, then this epoch allows shared read access from many different immutable references and owner. Otherwise, this epoch allows only a singular owner or mutable reference (not both) to read or write data. So we can regard the modification to the same data in the same epoch as one single modification, and \tool guarantees that at the end of each epoch, if the data is changed, the global address (including the tag) will be changed. And we can do this because in a single epoch if anyone modifies the data, the subsequent read to the data can only be done through that modifier which guarantees to have the most recent view of the data. So this Global-Address-Change-on-Write Property property holds after we regard the modifications in the same epoch as a single write operation.
The complexity in this invariant arises from the inclusion of the \textit{U} bit, which is employed as an optimization to reduce the frequency of color updates. As detailed at the end of \S~\ref{sec:owner}, every time an immutable reference is created through the owner or a mutable reference, the \textit{U} bit of that entity is reset to 0. Subsequently, when mutable dereferencing occurs on a mutable reference or \codeIn{Box} pointer, the color in that entity will be changed and the \textit{U} bit will be set to 1. Future mutable dereferencing through the same owner or mutable reference will not change color again until an immutable reference is borrowed from it. This optimization does not violate this \textit{Global-Address-Change-on-Write} invariant, because access to data in Rust's ownership model can be conceptually divided into epochs. Within each epoch, if immutable references exist, shared read access is permitted through multiple immutable references and the owner. Otherwise, the epoch allows only a single entity\textemdash either the owner or a mutable reference, but not both\textemdash to read or write data. Thus, modifications to the same data within the same epoch can be regarded as a singular modification. \tool ensures that at the end of each epoch, if the data has been altered, the colored global address will have changed. 
As for data accesses in a single epoch, if any entity modifies the data, subsequent reads can only occur through that modifier, which can directly read the updated data and does not involve the global address at all. Consequently, the \textit{Global-Address-Change-on-Write} invariant still holds with our optimization after considering modifications within the same epoch as a single write operation.

\subsection{Updated-Global-Address-Visible Invariant}

\vspace{.5em}
\begin{myquote} 
\begin{mdframed}
%\vspace{-.5em}
{\small
\setlength{\abovedisplayskip}{0pt}%
\underline{Updated-Global-Address-Visible Invariant}: All subsequently created shared immutable references can observe the updated global address corresponding to the latest write operation on the data object.
}
\end{mdframed}
\end{myquote}

% If it is created from a mutable reference, then clearly this mutable reference now holds the exclusive access rights to data so its global address field is the most recent version. If the data is modified through this mutable reference, then this global address field must have been updated. The immutable reference gets this global address and tag in store that in its own field. If it is created from the owner, then it is guaranteed by the ownership model that now no mutable reference exists to the same data. So if the data has ever been modified through any mutable references, they must have expired and written the updated global address and tag back to the owner, so the immutable reference can still gets this newest global address and tag.  

Immutable references are derived from either an owner, a mutable reference, or another immutable reference. When originating from a mutable reference, the latter maintains exclusive access to the data, ensuring its global address field reflects the most recent version. Any modifications made through this mutable reference necessitate updating the global address field. Consequently, subsequent immutable references inherit this updated colored global address.
In cases where an immutable reference originates from an owner, the ownership model ensures no concurrent mutable references exist for the same data. Thus, if the data has been modified through any previous mutable references, those references would have expired and updated the colored global address back to the owner. As a result, the newly created immutable reference acquires the latest colored global address. 
For immutable references derived from other immutable references, data mutation is prohibited during their existence. Hence, as long as the original immutable reference maintains the accurate colored global address, any new reference created from it will also possess the correct global address.

\subsection{Stale-Value-Elimination Invariant}

\vspace{.5em}
\begin{myquote} 
\begin{mdframed}
{\small
\setlength{\abovedisplayskip}{0pt}%
\underline{Stale-Value-Elimination Invariant}: Following a mutation of a value through an owner or a mutable reference, stale values stored in each server's read-only cache will not be accessed again.
}
\end{mdframed}
\end{myquote}

% This property can be proved from the address-change-on-write invariant and updated-address-visible invariant. As discussed in \S\ref{sec:acow-inv}, the access to a value can be divided into different epochs. In each epoch, the value can either be accessed through multiple immutable references or through a singular writer. The value can be cached on different servers in the shared immutable access epoch. Suppose now we are at the end of the N-th epoch of a value, and at the (N+1)-th epoch, the value is modified. Then the address-change-on-write invariant guarantees that the global address(including tag) of this value is changed. Then the next time the local cache hashmap is looked up through an immutable reference to this value, it must have the updated global address (according to the updated-address-visible invariant). This updated global address must be different from any previous global addresses of this value which are the keys in each local cache hashmaps. So no entries can be found, which means no stale values can be read from cache in \tool.

This invariant can be proved using the \textit{Global-Address-Change-on-Write} invariant and the \textit{Updated-Global-Address-Visible} invariant. As detailed in \S\ref{sec:acow-inv}, data access can be categorized into different epochs. During each epoch, a value is either accessible through multiple immutable references or a single mutable reference. The value may be cached on various servers in epochs of shared immutable access. Consider a scenario where we are at the end of the N-th epoch, and the value undergoes modification in the subsequent (N+1)-th epoch. The \textit{Global-Address-Change-on-Write} invariant ensures that the colored global address of this value is updated. Consequently, when the local cache hashmap is next queried in future epochs via an immutable reference to this value, it will reflect the updated global address, as dictated by the \textit{Updated-Global-Address-Visible} invariant. This new colored global address will differ from any prior global addresses associated with this value, which serve as keys in each server's local cache hashmaps. Therefore, no matching entries will be found in the hashmap, effectively preventing the reading of stale values in \tool.

Finally, the validity of the \textit{Data-Value} invariant is established through a proof by contradiction. If the \textit{Data-Value} invariant were violated, it would imply that a stale value is read post-mutation. The stale values only exist in the local read-only cache of each server, and values can only be mutated by the owner or mutable references. This indicates that a stale value copy stored in a server's read-only cache is accessed again after the value is mutated through the owner or mutable references, which contradicts the \textit{Stale-Value-Elimination} invariant. Hence, the \textit{Data-Value} invariant holds true for \tool.

By establishing both the \textit{SWMR} and \textit{Data-Value} invariants for \tool, we have proved the correctness of our algorithm as well as the memory coherence of \tool.

\section{Special Cases}
% \subsection{Tag Overflow}

% The 16-bit limitation of the tag field, though a rare occurrence, presents a possibility of overflow if an excessive number of references are continuously created on the same server. To address this, we implement a \textit{move-on-overflow} strategy. This method involves moving the value to a different address and resetting the tag to 0. Our approach ensures the uniqueness of each updated global address upon modifications to the data even when tag overflow happens.

% \mhr{When deallocating an object, we need to invalidate all cached copies of this address on all servers because this can potentially lead to reuse of this address on new objects. }

\subsection{Borrowing Stack Values and Partial Borrowing}

% In Rust, stack values can also be borrowed. A part of a struct can also be borrowed. So there exist mutable references to stack values and a part of a struct. These two scenarios are similar to each other because the problem is that their global address cannot be changed when accessing them through a mutable reference on a different server. So to solve this problem, we employ a copy and write-back method. We copy the stack value or the part of a heap struct to local memory. And upon expiration of the mutable reference, we write back the modified local copies to the original memory location on the other server. Besides, \tool also atomically increments the tag field of the owner of the struct it partially borrows from. For stack values, since their is no \codeIn{Box} pointer owner, we adopts a eager eviction policy for immutable references to stack values. When the reference count is 0 in line of algorithm~\ref{alg:imm}, we eagerly delete the local cache copy and entry in that server. In this way, we guarantee the next time we encounter the reference to the same stack value, we will read from its original location. 
Rust allows for the borrowing of stack values and partial borrowing of struct components. This leads to scenarios where mutable references may exist to stack values or a part of a struct. The challenge in these cases is that their global addresses cannot be altered when accessed via a mutable reference on a different server. To resolve this, \tool employs a \textit{copy and write-back} method. It involves copying the stack value or a part of the heap struct to local memory. Upon the expiration of the mutable reference, the modified local copy is written back to the original memory location on the original server. Additionally, \tool atomically increments the color field of the struct owner from which a partial borrow occurs. For stack values, lacking a \codeIn{Box} pointer owner, an eager eviction policy is adopted for immutable references. When the reference count of a stack value drops to 0, we eagerly delete the local cache copy and its corresponding entry on that server, ensuring that subsequent references to the same stack value will read from its original location.

\subsection{Reference Creation and Ownership Transferring}
% When creating new shared immutable references from existing ones, only the original data address is duplicated, leaving the new reference's local copy field null. This approach ensures the integrity of the reference count in the hashmap and ascertains that the local copy address is genuinely local. Subsequent data accesses through this new reference prompt \tool to update the local copy field and the hashmap accordingly. 

% Similarly, when transferring exclusive access rights between servers through ownership transfer, \tool will deallocate any local copies present on the source server. This step is vital to maintain consistency and prevent memory leaks during ownership transitions in distributed environments.

During the creation of new shared immutable references from existing ones, only the global address field is duplicated, leaving the new reference's local copy field \codeIn{null}. This method maintains the integrity of the reference count in the hashmap and ensures that any local copy address is genuinely local. Following accesses through this new reference trigger updates to the local copy field and the hashmap by \tool.

In scenarios involving the transfer of ownership between servers, \tool deems it crucial to deallocate any local cache copies present on the source server. This step is essential to preserve consistency and prevent memory leaks during ownership transitions in distributed settings.

\newpage
\bibliographystyle{abbrv}
\bibliography{paper}
% \bibliographystyle{ACM-Reference-Format}

% \newpage
% \appendix

\end{document}